\documentclass[times, twocolumn]{aastex631}
\usepackage{graphicx, graphics}
\usepackage{CJK}
\usepackage{amsmath, amssymb, bm}
\newcommand{\uat}[2]{\href{http://astrothesaurus.org/uat/#1}{#2 (#1)}}

\begin{document}
\begin{CJK*}{UTF8}{gbsn}
\title{A Survey of Protoplanetary Disks Using the Keck/NIRC2 Vortex Coronagraph}

\shorttitle{Protoplanetary Disks Using the Vortex Coronagraph}

\author{Nicole L. Wallack}
\affiliation{Earth and Planets Laboratory, Carnegie Institution for Science, Washington, DC 20015, USA} 
% \email{nwallack@carnegiescience.edu}
\affiliation{Division of Geological \& Planetary Sciences, California Institute of Technology, MC 150-21, Pasadena, CA 91125, USA}

\author{Jean-Baptiste Ruffio}
\affiliation{Department of Astronomy \& Astrophysics,  University of California, San Diego, La Jolla, CA 92093, USA}

\author{Garreth Ruane}
\affiliation{Jet Propulsion Laboratory, California Institute of Technology, 4800 Oak Grove Drive, Pasadena, CA, 91109, USA}

\author{Bin B. Ren (任彬)}
\altaffiliation{Marie Sk\l odowska-Curie Fellow}
\affiliation{Universit\'{e} C\^{o}te d'Azur, Observatoire de la C\^{o}te d'Azur, CNRS, Laboratoire Lagrange, Bd de l'Observatoire, CS 34229, 06304 Nice cedex 4, France}
\affiliation{Universit\'{e} Grenoble Alpes, Institut de Plan\'{e}tologie et d'Astrophysique (IPAG), F-38000 Grenoble, France}
\affiliation{Department of Astronomy, California Institute of Technology, MC 249-17, 1200 East California Boulevard, Pasadena, CA 91125, USA}

\author{Jerry W. Xuan}
\affiliation{Department of Astronomy, California Institute of Technology, MC 249-17, 1200 East California Boulevard, Pasadena, CA 91125, USA}

\author{Marion Villenave}
\affiliation{Jet Propulsion Laboratory, California Institute of Technology, 4800 Oak Grove Drive, Pasadena, CA, 91109, USA}
\affiliation{Universit\'a degli Studi di Milano, Dipartimento di Fisica, via Celoria 16, 20133 Milano, Italy}

\author{Dimitri Mawet}
\affiliation{Department of Astronomy, California Institute of Technology, MC 249-17, 1200 East California Boulevard, Pasadena, CA 91125, USA}
\affiliation{Jet Propulsion Laboratory, California Institute of Technology, 4800 Oak Grove Drive, Pasadena, CA, 91109, USA}

\author{Karl Stapelfeldt}
\affiliation{Jet Propulsion Laboratory, California Institute of Technology, 4800 Oak Grove Drive, Pasadena, CA, 91109, USA}

\author{Jason J. Wang}
\affiliation{Department of Astronomy, California Institute of Technology, MC 249-17, 1200 East California Boulevard, Pasadena, CA 91125, USA}
\affiliation{Center for Interdisciplinary Exploration and Research in Astrophysics (CIERA) and Department of Physics and Astronomy, Northwestern University, Evanston, IL 60208, USA}

\author{Michael C. Liu}
\affiliation{Institute for Astronomy, University of Hawaii, 2680 Woodlawn Drive, Honolulu, HI 96822, USA}

\author{Olivier Absil}
\affiliation{STAR Institute, Universit\'e de Li\`ege, 19c All\'ee du Six Ao\^ut, 4000 Li\`ege, Belgium}

\author{Carlos Alvarez}
\affiliation{W. M. Keck Observatory, Kamuela, Hawaii, USA}

\author{Jaehan Bae}
\affiliation{Department of Astronomy, University of Florida, Gainesville, FL 32611, USA}
\affiliation{Earth and Planets Laboratory, Carnegie Institution for Science, Washington, DC 20015, USA}

\author{Charlotte Bond}
\affiliation{Science and Technology Facilities Council (STFC), Polaris House, North Star Avenue, Swindon, SN2 1SZ, United Kingdom}

\author{Michael Bottom}
\affiliation{Institute for Astronomy, University of Hawaii, 2680 Woodlawn Drive, Honolulu, HI 96822, USA}

\author{Benjamin Calvin}
\affiliation{Department of Astronomy, California Institute of Technology, MC 249-17, 1200 East California Boulevard, Pasadena, CA 91125, USA}
\affiliation{Department of Physics \& Astronomy, University of California Los Angeles, Los Angeles, CA 90095, USA}

\author{\'Elodie Choquet}
\affiliation{Aix Marseille Univ, CNRS, CNES, LAM, Marseille, France}

\author{Valentin Christiaens}
\affiliation{STAR Institute, Universit\'e de Li\`ege, 19c All\'ee du Six Ao\^ut, 4000 Li\`ege, Belgium}

\author{Therese Cook}
\affiliation{Department of Physics \& Astronomy, University of California Los Angeles, Los Angeles, CA 90095, USA}

\author{Bruno Femen\'ia Castell\'a}
\affiliation{Instituto de Astrof\'isica de Canarias, C/ V\'ia L\'actea s/n, E-38205 San Crist\'obal de la Laguna, Spain}
\affiliation{ Departamento de Astrof\'isica, Universidad de La Laguna, E-38205 San Crist\'obal de la Laguna, Spain}

\author{Carlos Gomez Gonzalez}
\affiliation{Centre Nacional de Supercomputaci\'o, Pla\c{c}a d'Eusebi G\"uell, 1-3, 08034 Barcelona, Spain}

\author{Greta Guidi}
\affiliation{Institut de RadioAstronomie Millim\'etrique (IRAM), 300 rue de la Piscine, 38406 Saint Martin d'H\'eres }

\author{Elsa Huby}
\affiliation{LESIA, Observatoire de Paris, Universit\'e PSL, CNRS, Sorbonne Universit\'e, Universit\'e de Paris Cit\'e, 5 place Jules Janssen, 92195 Meudon, France}

\author{Joel Kastner}
\affiliation{Center for Imaging Science, Rochester Institute of Technology, Rochester, NY 14623, USA}
\affiliation{School of Physics and Astronomy, Rochester Institute of Technology, Rochester, NY 14623, USA}
\affiliation{Laboratory for Multiwavelength Astrophysics, Rochester Institute of Technology, Rochester, NY 14623, USA}

\author{Heather A. Knutson}
\affiliation{Division of Geological \& Planetary Sciences, California Institute of Technology, MC 150-21, Pasadena, CA 91125, USA}

\author{Tiffany Meshkat}
\affiliation{IPAC, California Institute of Technology, M/C 100-22, 1200 East California Boulevard, Pasadena, CA 91125, USA}

\author{Henry Ngo}
\affiliation{NRC Herzberg Astronomy and Astrophysics, 5071 West Saanich Road, Victoria, British Columbia, Canada}

\author{Sam Ragland}
\affiliation{W. M. Keck Observatory, Kamuela, Hawaii, USA}

\author{Maddalena Reggiani}
\affiliation{Institute of Astronomy, KU Leuven, Celestijnenlaan 200D, B-3001 Leuven,
Belgium}

\author{Luca Ricci}
\affiliation{Department of Physics and Astronomy, California State University Northridge, 18111 Nordhoff Street, Northridge, CA 91130, USA}

\author{Eugene Serabyn}
\affiliation{Jet Propulsion Laboratory, California Institute of Technology, 4800 Oak Grove Drive, Pasadena, CA, 91109, USA}

\author{Taichi Uyama}
\affiliation{IPAC, California Institute of Technology, M/C 100-22, 1200 East California Boulevard, Pasadena, CA 91125, USA}

\author{Jonathan P. Williams}
\affiliation{Institute for Astronomy, University of Hawaii, 2680 Woodlawn Drive, Honolulu, HI 96822, USA}

\author{Peter Wizinowich}
\affiliation{W. M. Keck Observatory, Kamuela, Hawaii, USA}

\author{Zoe Zawol}
\affiliation{Department of Astronomy, California Institute of Technology, MC 249-17, 1200 East California Boulevard, Pasadena, CA 91125, USA}

\author{Shangjia Zhang}
\author{Zhaohuan Zhu}
\affiliation{Nevada Center for Astrophysics, University of Nevada, Las Vegas, 4505 S. Maryland Parkway, Las Vegas, NV 89154, USA}
\affiliation{Department of Physics and Astronomy, University of Nevada, Las Vegas, 4505 S. Maryland Parkway, Las Vegas, NV 89154, USA}

\shortauthors{Wallack et al.}

\begin{abstract}
Recent Atacama Large Millimeter/submillimeter Array (ALMA) observations of protoplanetary disks in the millimeter continuum have shown a variety of radial gaps, cavities, and spiral features. These substructures may be signposts for ongoing planet formation, and therefore these systems are promising targets for direct imaging planet searches in the near-infrared. To this end, we present results from a deep imaging survey in the $L'$-band (3.8 $\mu$m) with the Keck/NIRC2 vortex coronagraph to search for young planets in 43 disks with resolved features in the millimeter continuum or evidence for gaps/central cavities from their spectral energy distributions. Although we do not detect any new point sources, using the vortex coronagraph allows for high sensitivity to faint sources at small angular separations (down to ${\sim}0\farcs1$), allowing us to place strong upper limits on the masses of potential gas giant planets. We compare our mass sensitivities to the masses of planets derived using ALMA observations, and while we are sensitive to $\sim$1~M$_{\tt Jup}$ planets in the gaps in some of our systems, we are generally not sensitive to planets of the masses expected from the ALMA observations. In addition to placing upper limits on the masses of gas giant planets that could be interacting with the dust in the disks to form the observed millimeter substructures, we are also able to map the micron-sized dust as seen in scattered light for 8 of these systems. Our large sample of systems also allows us to investigate limits on planetary accretion rates and disk viscosities. 
\end{abstract}

\section{Introduction}
Gaps, cavities, and spiral features seen in protoplanetary disks are thought to be the signposts of planet formation. Millimeter-sized dust grains are expected to rapidly drift inwards, resulting in the depletion of large dust grains at large radii \citep{Weidenschilling1977}. However, when imaged by facilities such as the Atacama Large Millimeter/submillimeter Array (ALMA), structures in the large dust grains are evident out to hundreds of astronomical units in many protoplanetary disks (e.g., \citealt{Andrews2018, Long2018}). In order to keep these dust grains at such large orbital separations, there must be mechanisms in effect that trap dust grains at large distances and produce the spiral, concentric ring, and gap substructures that seem to be prevalent among protoplanetary disks imaged with ALMA. A number of different mechanisms have been suggested to explain the protoplanetary disk dust distributions, including magnetorotational instabilities, dead zones, and condensation fronts (e.g., \citealt{Isella2016, Liu2018, Ohashi2019}). However, one of the most favored explanations for the substructures in the observed millimeter dust in these disks is that they are caused by forming planets interacting with the natal disk structure. Forming giant planets can cause pressure bumps outside of their orbits and trap large dust grains into rings while less massive planets generally open gaps in the dust in the disks with no significant changes to the local gas \citep{Paardekooper2006, Zhu2014}. The presence of protoplanets associated with disk substructures has also been inferred by measuring deviations from the Keplerian velocity in the gas velocity field (\citealt{Pinte2019,Teague2018}).

Determining the masses of the planets that are carving out the gaps in the gas and dust of the disk is complicated by uncertainties on the bulk disk parameters, which can lead to large degeneracies in the masses derived from the geometries of millimeter dust cavities and gaps. For example, assuming different values for the disk viscosity, which is a largely observationally unconstrained parameter, can lead to differences in the calculated masses of planets on the order of a factor of  $>$4 \citep{Zhang2018}. One way to more directly estimate the masses of planets carving these gaps is to directly detect the young planets' thermal emission. 

Direct imaging is uniquely able among planet detection and characterization methods to probe the locations of the proposed wide-separation planets carving out the gaps in these young systems. Thus, direct imaging is capable of detecting planets in the act of forming, allowing us to place observational constraints on the masses required to carve out millimeter gaps and also on formation locations and accretion timescales. Leveraging the fact that both the blackbody expectation and predictions from more rigorous atmospheric models \citep{Skemer2014} of a cool planet would dictate more advantageous contrast in the $L'$ band ($3.8~\mu$m) over other wavelengths, there have been multiple direct imaging surveys of protoplanetary disks at $L'$ (e.g., \citealt{Stone2018,Launhardt2020, Cugno2023}) and more broadly in the infrared \citep{Asensio-Torres2021}. Moreover, observing in the $L'$ band allows for a compromise between more favorable planet-to-star contrast expectations and increased background noise at longer infrared wavelengths such as the $M$ band. 

Conducting a survey at $L'$ also allows for more advantageous planet-to-star flux ratios over millimeter wavelengths making these infrared surveys the most sensitive to directly detecting the planets inferred by the millimeter data. Additionally, infrared wavelength observations also  provide information on the micron-sized dust distribution as inferred from scattered light. This means that this technique allows for both possible direct detection of planets while also providing a better understanding of the distribution of small dust grains, allowing for direct observations of planets interacting with their natal disks and actively accreting their atmospheres (e.g., the PDS 70 system; \citealt{Benisty2018, Keppler2018}).

To this end, we present a survey using the Keck/NIRC2 vortex coronagraph (\citealt{Mawet2016, Serabyn2017}) of protoplanetary disks previously observed at other wavelengths that show substructures and/or cavities. Utilizing the vortex coronagraph allows for high sensitivity to faint sources at small angular separations (down to $0\farcs1$), ideal for probing the cavities shown in the millimeter/submillimeter dust close to the host stars. 

However, owing to our sample of systems being comprised of those with known disk structure, an important caveat to our sensitivity to planets is that the effects of extinction from the circumstellar disks can also lead to uncertainties in the derived planet masses. Using hydrodynamic simulations, \cite{sanchis2020} found extinction coefficients, A$_{L}$, for a 2 M$_{\tt Jup}$ planet at 100 au to be 1.31 and for a 5 M$_{\tt Jup}$ planet at the same location to be 0.02. Moreover, determining the exact level of the extinction caused by the circumstellar disk in these systems is nontrivial and indeed the values of the extinction are dependent on the surface densities of the gas and dust \citep{Cugno2023}. Therefore, the masses derived from direct imaging in the infrared, while more directly estimated than using the geometries of the dust and gas at millimeter wavelengths, come with the large caveat that extinction, and the uncertainties on extinction, may alter the derived values.

In Section ~\ref{sec:observationStrategy} we discuss the strategy for our observations. In Section ~\ref{sec:Results}, we present detection limits of all our targets using our derived contrast curves and in Section ~\ref{sec:Discussion} we put these contrast curves into the broader context of the detection possibility of long-period giant planets.

\section{Observations}\label{sec:observationStrategy}

\subsection{Target Selection}
All targets were selected for using the criteria that they (1) have published images in the millimeter/submillimeter or a published spectral energy distribution (SED); (2) show evidence in the ALMA data or SED for a gap and/or cavity that could have been formed due to interactions between the millimeter-sized dust and a forming protoplanet; (3) are nearby (preferably within ${\sim}300$~pc); (4) are bright enough for a high-quality adaptive optics (AO) correction; and (5) are estimated to be young ($<$30 Myr). Our targets were originally taken from papers on individual disks observed with ALMA and disks that showed evidence of gaps from their SEDs. However, following the publication of the ALMA survey of the Taurus molecular cloud \citep{Long2018}, and the DSHARP survey \citep{Andrews2018}, we mostly selected targets that were from one of these two surveys because their high-resolution observations allowed for more precise estimates of gap locations in the disks and expected masses of planets that could be opening those gaps (see Table~\ref{table:systems} for more information).

\subsection{Observing Strategy}
Observations started in observing semester 2015B and continued over the next 10 observing semesters (see Table~\ref{table:observations} for more details). We observed a total of 43 protoplanetary disk targets, with some observed multiple times for follow-up observations or to account for non-ideal observing conditions. 

Observations were done using two different wavefront sensors (WFSs). The observations done prior to 2019B (2019 August 1) were done primarily with the visible Shack-Hartmann WFS on the Keck~II telescope. The AO correction for this WFS is operated in the $R$ band, meaning that early in our survey we were limited to targets with bright $R$ band magnitudes. For the post-2019B targets in our survey, we used the Keck~II infrared pyramid WFS, which performs wavefront sensing in the $H$ band, meaning that we could observe redder targets with a better AO correction \citep{Bond2020}. We observed targets down to $H$ = 11.9 and $R$ = 14.39. Some targets were first observed using the Shack-Hartmann WFS and subsequently observed with the infrared pyramid WFS in order to achieve a better AO correction.

Our targets were observed with the Keck/NIRC2 vortex coronagraph using \texttt{QACITS}, a real-time point spread function (PSF) centering algorithm that keeps the target star well centered on the vortex mask and stabilized during the entirety of the observing sequence \citep{Huby2017}. The vortex coronagraph allows for high-contrast imaging at small angular separations (down to ${\sim}100$~mas) in the $K$, $L'$, and $M$ bands (see \citealp{xuan2018} for a performance characterization of the vortex coronagraph). 

A combination of using the vortex coronagraph and multiple observing techniques (angular differential imaging (ADI) and reference star differential imaging (RDI)) allows for the most advantageous scenario for the detection of exoplanets close in to their host stars. ADI and RDI are largely complementary observing strategies enabling the detection of point sources close to the central star and the characterization of any extended structure in the image. 

The ADI strategy \citep{Liu2004, Marois2006} takes advantage of the field rotation of the sky as seen by alt-azimuth telescopes, causing any circumstellar sources (planet or disk) to rotate with time on the detector and any non-astrophysical signals due to optical effects (such as diffraction and high-order wavefront errors) to stay mostly static in time. After subtracting most quasi-static speckles using the sequence of frames, the astrophysical sources should remain, allowing for possible detection of disk structures or point sources that would otherwise be embedded in the diffracted starlight from the host star. ADI is most sensitive to point sources at regions a few diffracted beam widths ($\lambda/$D, where $\lambda$ is the wavelength and $D$ is the telescope diameter) from the star. 

The RDI observing strategy uses the PSF of similar stars (in terms of stellar type and airmass at the time of observation) to characterize and remove any remaining stellar contribution (which was not suppressed by the coronagraph). See \citealt{Ruane2019} for a full discussion of the benefits of utilizing RDI in the context of observing with the vortex coronagraph on Keck$/$NIRC2. RDI is most sensitive to areas at smaller angular separations (${\lesssim}0\farcs$3) in the frame, making it beneficial to use both observing strategies \citep{Ruane2017}. We utilize two methods of RDI, targeted RDI and single-night RDI. While we observed all of our targets in the $L'$ band at least using ADI, for a subset of the targets we also obtained specifically targeted reference stars for RDI. These targeted RDI observations consisted of observing our science target symmetrically around transit (minimum airmass) and then observing a specifically chosen reference star (to be as similar in magnitude and airmass as possible) for the same amount of time also centered at its transit. We chose reference stars to best represent the PSFs of our science targets but without known planets, stellar companions, or disk material. When possible, we observed each reference star centered on its time of transit for enough time so that the total number of frames and integration time were similar to those of the science target. Our reference stars were selected to match the AO performance between the reference and science targets. Therefore, we matched the $H$-band magnitudes when using the infrared pyramid WFS and $R$-band magnitudes when using the Shack-Hartmann WFS.

In addition to the use of specifically targeted reference stars, we also utilized RDI within an observing night when multiple targets were observed at $L'$ within a night. Despite the stellar PSFs not being manually matched to those of the science targets, in the absence of specifically selected reference stars, this form of single-night RDI gives us a preliminary analysis that often achieves better contrast at smaller separations ($\lesssim 0\farcs$3) than ADI. A more refined RDI analysis with optimized reference frames taken from the library of $L'$ vortex observations might allow for deeper achieved contrasts. 

Regardless of the observing strategy, we corrected the frames for bad pixels, flat-fielded them, subtracted the sky background, registered them, and applied principal component analysis (PCA). PCA was utilized in order to estimate the stellar contribution in the images via the Vortex Image Processing (\texttt{VIP}) package\footnote{Available under open-source license at \url{https://github.com/vortex-exoplanet/VIP}.} \citep{Gonzalez2017} and the \texttt{pyKLIP} package\footnote{Available under open-source license at \url{https://bitbucket.org/pyKLIP/pyklip}.} \citep{Wang2015}.

\section{Results}\label{sec:Results}
\subsection{Survey Sensitivity Estimates}
\begin{figure}
\includegraphics[width=0.44\textwidth]{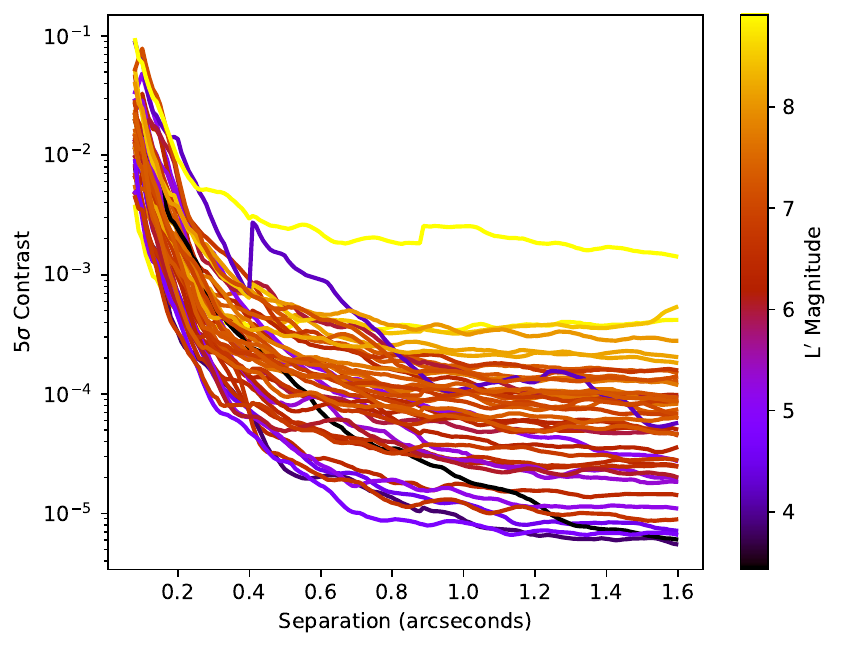}
{\caption {Optimal contrast curves for our sample of protoplanetary disks colored by the $L'$ magnitude. Each line is the most optimal 5$\sigma$ contrast for a different disk. The contrast curves at select separations are shown in Table~\ref{table:contrasts} in the Appendix.}
\label{figure:cc}}
\end{figure}

We first compute the 5$\sigma$ contrast curves for each of our targets using \texttt{VIP} (Figure~\ref{figure:cc}). These contrast curves are computed in the same manner as described in \cite{xuan2018}. In brief, for each observation, we determine the optimal contrast curve among our different combinations of inner and outer mask sizes, reduction method (ADI and RDI), and numbers of principal components by determining the optimal contrast in any of the different combinations of those parameters in steps of 1 pixel. We do this by first generating a 5$\sigma$ contrast curve for each combination of mask and frame size, number of principal components, and reduction method using \texttt{VIP} \citep{Gonzalez2017} and accounting for the effects of small sample statistics as described in \cite{Mawet2014}. When accounting for the small sample statistics, we fix the false positive fraction to 2.867 $\times$ 10$^{-7}$, which is equivalent to using 5$\sigma$ in the Gaussian case. We call the contrast curve generated using this assumption our 5$\sigma$ contrast. \texttt{VIP} utilizes the fake companion method of determining the throughput \citep{Gonzalez2017}. This method involves the injection and recovery of a series of fake companions into the images in order to determine the contrast limits. We generate a 5$\sigma$ contrast curve for each combination of inner and outer mask size, reduction method, and number of principal components and then compare the contrast at each 1 pixel step obtained for all versions of the contrast and take the minimum contrast achieved. We repeat this process for each 1 pixel step until we have an optimal contrast curve across all radial separations, which in reality represents multiple post-processing frame configurations.  In the case that we have multiple datasets of the same object, we only consider the contrast curve corresponding to the more sensitive contrast in subsequent analyses (and present the optimal contrast curve for each system in Figure~\ref{figure:cc} and the corresponding contrast curves at select separations in Table~\ref{table:contrasts}). 

\begin{figure}
{\includegraphics[width=0.44\textwidth]{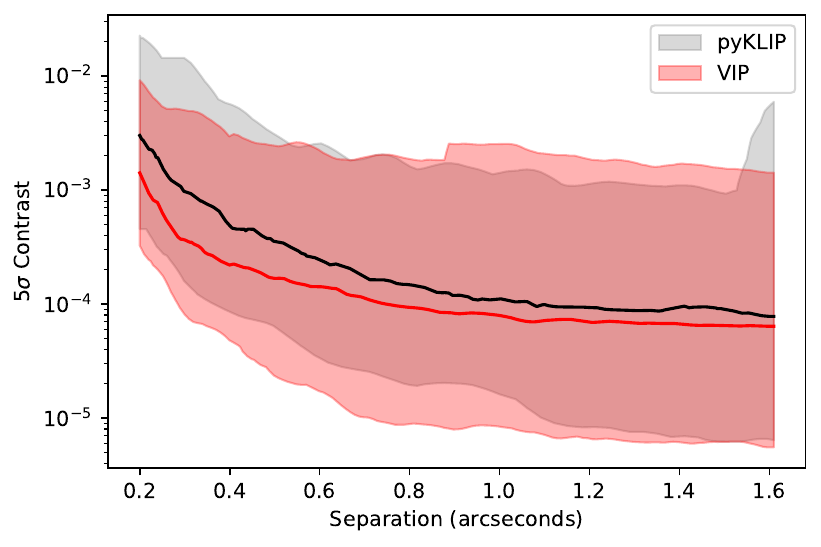}}
{\caption {Comparison of the contrasts achieved using the \texttt{pyKLIP} and \texttt{VIP} reductions. The solid lines are the median contrasts of all of the contrast curves and the shaded regions represent the range of contrasts. We use ADI for the \texttt{pyKLIP} reductions and a combination of ADI and RDI for the \texttt{VIP} reductions. The improved performance of the \texttt{VIP} reductions as compared to the \texttt{pyKLIP} reductions is likely due to the additional observation-specific optimizations that were done using \texttt{VIP}. However, the contrasts are still comparable, showing the robustness of our calculated contrasts.}
\label{figure:comparison}}
\end{figure}

For our study, we focus primarily on our \texttt{VIP} reductions, but we also compute the corresponding contrast curves using the \texttt{pyKLIP} package. After image preprocessing and registration, each image is high-pass filtered using a Gaussian kernel with a width equal to twice the FWHM of the off-axis (when the star is not on the vortex coronagraph) PSF. We then run \texttt{pyKLIP} in ADI mode with the following parameters: an exclusion parameter of half the FWHM, a maximum of 30 Karhunen--Lo\`eve modes, an inner working angle equal to twice the FWHM, an annuli width equal to about half the FWHM, and a number of reference frames limited to a maximum of the 200 most correlated images for each science frame \citep[see][for a more detailed description of the algorithm]{Ruffio2017}. After speckle subtraction, the images of a dataset are coadded and a Gaussian matched filter is used as described in \citet{Ruffio2017}. The width of the Gaussian kernel is equal to the PSF FWHM. The algorithm throughput is calculated from the injection and recovery of 160 simulated planets. Contrast curves are computed by taking the standard deviation of concentric annuli of the throughput-corrected match filtered images taking into account small sample statistics. Despite high-pass filtering the \texttt{pyKLIP} reductions and not high-pass filtering the \texttt{VIP} reductions, the contrasts that are achieved by the two reductions are comparable (Figure~\ref{figure:comparison}). We utilize both sets of reductions in order to thoroughly vet detected point sources. For  \texttt{pyKLIP}, we primarily focus on the aforementioned ADI reductions. However, with \texttt{VIP}, we utilize a combination of ADI and RDI, because RDI outperformed the ADI at small separations using \texttt{VIP}, which likely helps to account for the better contrast limits for the \texttt{VIP} reductions at small separations. Additionally, the use of observation specific optimizations using \texttt{VIP} could also explain the improved performance of the \texttt{VIP} reductions.

\subsection{Point Source Detection}\label{point_sources}
In order to determine if we detect any significant point sources in our images, we generate a signal-to-noise ratio (S/N) map for each different combination of frame sizes, number of principal components, and reduction methods that were determined to be the optimal combination at each 1 pixel step in separation utilizing the \texttt{VIP} package and doing a correction to  account for the effects of small sample statistics according to \cite{Mawet2014}. We then create an optimal S/N map representative of our composite optimal contrast curve by taking the S/N map corresponding to the frame size, number of principal components, and reduction method that best optimizes the contrast in each 1 pixel step and making a composite S/N map consisting of each of the respective 1 pixel annuli. Since we are interested in the regions of the disks where structure is seen in the millimeter, we focus on the region between 0$\farcs$15 and 1$\farcs$6.

Direct imaging data are plagued by stellar residuals due to imperfect removal of the stellar PSFs. Some of these residual speckles can mimic the appearance of point sources in our data. Without the use of multiwavelength data of these speckles in order to ascertain whether they have spectra similar to that of the host star, these speckles can be mistaken for astrophysical point sources. However, many of these speckles are post-processing specific and different reduction pipelines can generate artifacts in different locations and at different S/Ns.

We determine if there are any physical point sources detected in any of our observations by first generating a list of all point sources outside of our central $0\farcs15$ mask and inside $1\farcs6$ with S/N $>5$, utilizing the S/N maps generated from the \texttt{VIP} reductions (the locations of which are shown in Figure~\ref{figure:blobs}). We achieve slightly better contrast limits using the \texttt{VIP} reductions (Figure~\ref{figure:comparison}), so we search for point sources in that set of reductions. We exclude the known binary of AS 205 from our list of detected point sources in order to search for previously unknown companions. While we also do clearly detect PDS~70b in our observations, we do not discuss the planet nor its disk (other than showing the accompanying scattered light disk in the context of its millimeter continuum in Section~\ref{disks}), as we have thoroughly characterized the planet and its disk using the data presented herein in \cite{Wang202070}. 

\begin{figure}
{\includegraphics[width=.44\textwidth]{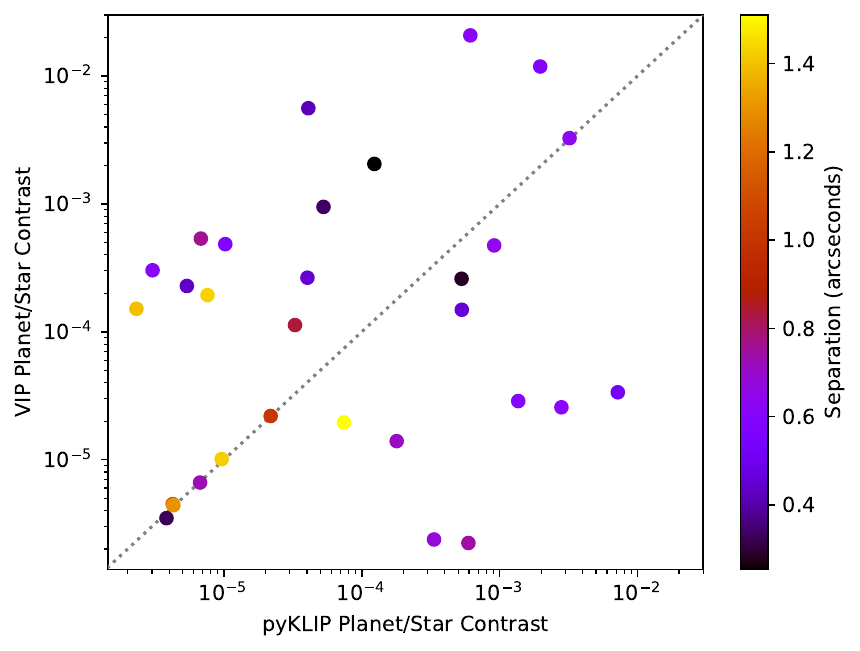}}
{\caption {Comparison of the planet-to-star flux ratio of the point sources in Figure~\ref{figure:blobs} determined from the \texttt{VIP} reductions vs. the planet-to-star flux ratio in the same location in the \texttt{pyKLIP} reductions. A perfectly consistent flux ratio is designated by the dotted line.}
\label{figure:blob_comparison}}
\end{figure}

In order to determine whether the detected point sources are real astrophysical signals instead of residual diffracted starlight, we compare the contrasts of the point sources detected with the \texttt{VIP} reductions to those of the contrasts at the same locations in the observations analyzed using \texttt{pyKLIP} (Figure~\ref{figure:blob_comparison}). If these point sources were real, the fluxes measured between the two reductions should be consistent. The S/N maps using \texttt{pyKLIP} are derived from computing the standard deviation in 2 pixel wide concentric annuli. For each of the 30 point sources with S/N $>5$ in the \texttt{VIP} S/N maps, we determine the flux and uncertainty of that point source using the maximum likelihood estimation method of the negative fake companion technique implemented in \texttt{VIP} for both our \texttt{VIP} and \texttt{pyKLIP} reductions.

The majority (24) of the 30 point sources detected in the \texttt{VIP} reductions do not have consistent (at the 2$\sigma$ level or better) fluxes detected in the \texttt{pyKLIP} reductions and are therefore likely not physical. The other six point sources represent a combination of real disk detections that are consistent between the different reductions and observations where there are many residual speckles in the datasets due to imperfect speckle subtraction by both reduction algorithms. Of these six point sources, one (in  HD~179218) is due to residual effects in the \texttt{VIP} RDI reductions due to poor reference star PSFs and a particularly noisy corresponding \texttt{pyKLIP} ADI reduction, two (one in CQ~Tau and one in LkCa~15) are due to real disk structures that appear to be point-like but are consistent with extended structure, two (in HL~Tau) do not have the typical PSFs of point sources and instead have extended appearances likely due to diffraction from the telescope support structures, and one appears in the AS 209 dataset with the poorer contrast and not in the observation with a better detection limit, indicating that it is not real. An additional way of assessing the validity of possible point sources is whether the planet-to-star contrast of a point source as determined from the S/N maps would have been detected given our contrast curves. While we are vetting the point sources with S/N $>5$, this does not necessarily mean that we would be sensitive to a source of that contrast as determined from our contrast curves. Only points with contrasts above our sensitivity would have been able to be detected. Indeed, only 14 of our point sources even would have been properly detected (i.e. having contrasts above our detection sensitivity), and of these 14, none of them have consistent fluxes (at the 2$\sigma$ level or better) between the two reductions.

The appearance of disk features being post-processed into apparent point-sources is a well-known phenomenon \citep{Currie2019}, and indeed the presence of protoplanet(s) in LkCa 15 has been debated (e.g., \citealt{Currie2019, Kraus2012, Follette2023}). Without detailed disk modelling, it is nontrivial to disentangle disk material from a point source, therefore to err on the side of caution in the absence of such modelling (and because the flux levels of our two reductions did not agree for these points likely due to differences in post processing), we elect to say that we do not convincingly detect planets that are coincident with the disk material in CQ~Tau and LkCa~15. Therefore, by a combination of automatic and manual vetting, we do not convincingly detect any new point sources in our observations in the area between 0$\farcs$15 and 1$\farcs$6.

While we utilized both \texttt{pyKLIP} and \texttt{VIP} in order to vet potential point sources in our data, our \texttt{VIP} reductions generally achieved better contrasts than our \texttt{pyKLIP} reductions (Figure~\ref{figure:comparison}). Therefore, in all subsequent analyses presented herein, we present results utilizing our \texttt{VIP} reductions.

\subsection{Scattered Light Disks}\label{disks}
\begin{figure*}
\begin{centering}
\includegraphics[width=\textwidth]{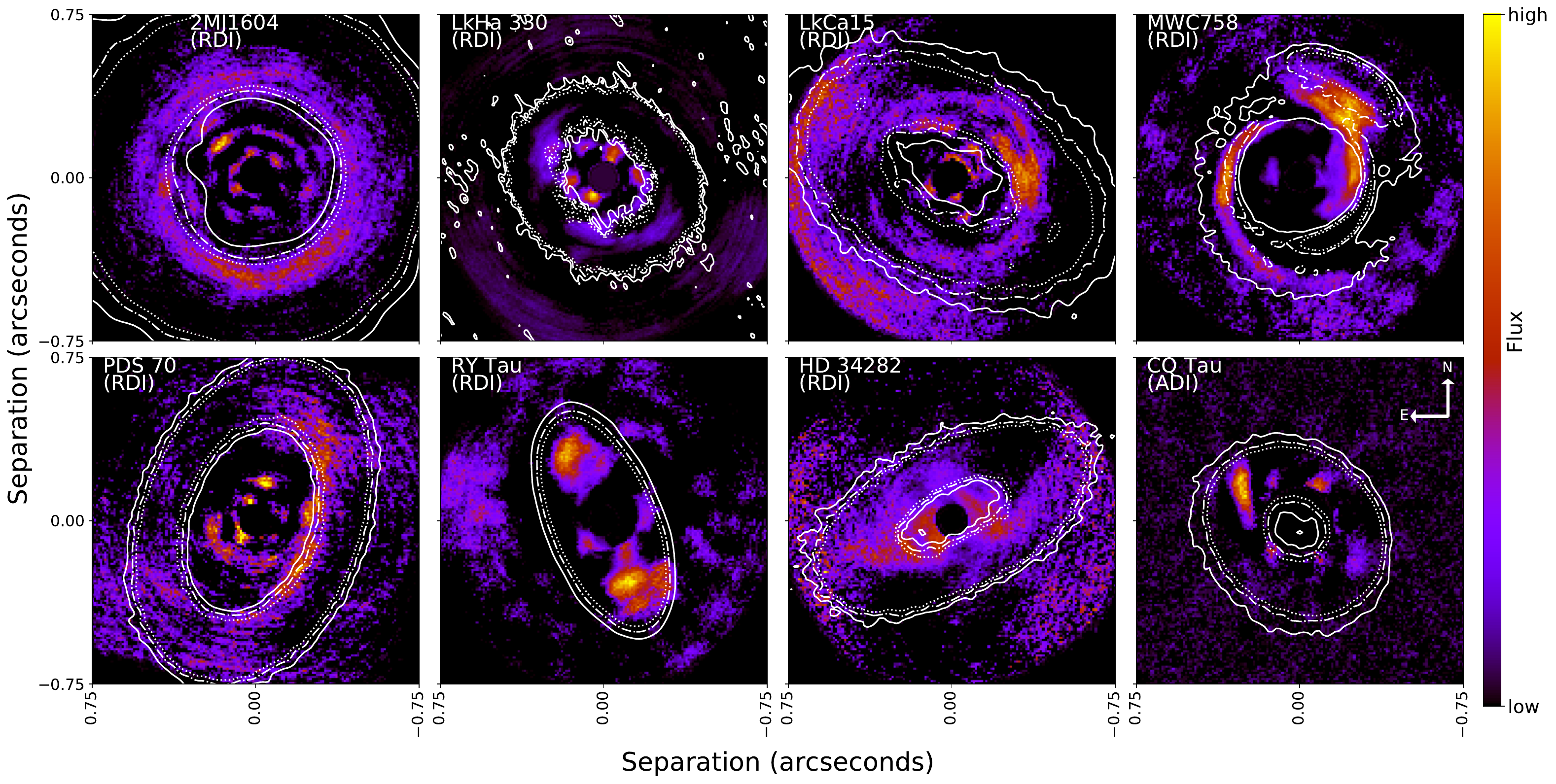}
\caption{Disks in our survey showing evidence of structure which is indicative of scattered light disks. We show the 1$\sigma$, 2$\sigma$, and 3$\sigma$ ALMA contours (from references in Table~\ref{table:systems}) overlaid on our 3.8 micron images for reference. North is up and east is left in all of the images. We show the reduction method that optimizes the appearance of the extended structure.} 
\label{figure:disks}   
\end{centering}
\end{figure*}

The direct detection of substructures in large grain dust as detected by ALMA and/or evidence of a gap or cavity in the stellar SED was a requirement of our target selection. Therefore, we might expect to see evidence of disk structure in all of our observations. However, in the case of the ALMA observed disks, the thermal emission of the millimeter/submillimeter sized dust grains is what is being detected, whereas in the case of near-IR observations, we are sensitive to the scattered light of the micron-sized dust. Owing to viewing geometries, the scattering properties of dust grains at different wavelengths, the luminosity of the host star, data reduction techniques, and the sensitivity of our study, we would not expect to see evidence of scattered light disks in all of our targets. We do, however, see disk structures in a subset of our observations which are consistent with the ALMA observations for these systems (Figure~\ref{figure:disks}). We optimize the images shown in Figure~\ref{figure:disks} by selecting the number of principal components in our full frame images that best shows the extended structure. We reserve the direct modeling of these disks and the discussion related to the small grain dust distribution to a forthcoming study.

For purposes of our initial survey, we limit our RDI analyses to single nights (where to first order, we expect the most similar AO correction) where the stellar PSFs of any other objects observed on the same night with the vortex coronagraph at $L'$ act as our reference library. In this initial census of detected scattered light disks, we do not optimize this reference library herein, therefore we are not detecting every disk that would be evident from a library where stellar properties are more closely matched to the science target (such as the SR~21 disk that we present in \citealp{Uyama2020EarlyDisk}, where the reference PSFs are selected across different observing nights). We also do not combine multiple observing nights that might allow for the increase in S/N needed to make the scattered light disk more evident (such as in the case of the HD~163296 disk presented in \citealp{Guidi2018High-contrastCoronograph}, where the disk was more visible when data over two consecutive nights were combined). 

\subsection{Stellar Parameters}\label{stellar_parameters}
Planet mass limits derived from observational contrast limits are a strong function of age. Therefore it is of the utmost importance that we determine accurate and self-consistently derived ages for our host stars (and therefore planetary systems). However, the ages of young stars are notoriously difficult to determine, so in order to allow for direct, population-level analyses of planet mass upper limits, we derive uniform ages for our sample of systems.

In order to estimate the stellar ages for these protoplanetary disk-hosting stars, we first attempt to compile the stellar parameters ($\rm L_\star$, $\rm T_{eff}$) required to calculate the ages of these systems as consistently as possible. Since many of the stellar parameters are sourced from references that predate updated Gaia distances \citep{GaiaCollaboration2021Special3}, when applicable, we first rescale the stellar luminosities from the sources in Table~\ref{table:systems} to account for the updated distances from Gaia. In the cases where uncertainties on the stellar luminosity and effective temperature are not published, we assume a 10\% uncertainty in log($\rm L_\star$) and 2\% uncertainty in $\rm T_{eff}$ \citep{Pascucci2016}.

We then estimate the stellar ages and masses using isochrones from \cite{Choi2016} in the range of 0.5 to 50~Myr. We interpolate the tracks to probe the mass range from 0.05 to 1.4~Msun, in steps of 0.01~Msun. We adopt the method described in \cite{Andrews2013TheHosts} to determine stellar masses and ages. We first evaluate a likelihood function for each set of luminosities and effective temperatures \citep[equation 1 in][]{Andrews2013TheHosts}. We then marginalize the likelihood distribution and take the median of the marginalized distribution to estimate the age and mass of the star. We estimate the uncertainties as the 16\% and 84\% percentiles of these marginalized distributions (which takes into account the uncertainties on the stellar luminosity and effective temperature). 

Our calculated ages are generally consistent at the 1$\sigma$ level with those presented in the references in Table~\ref{table:systems}, with variations in the ages mostly due to updates to the distances to these stars as determined from updated parallaxes from Gaia \citep{GaiaCollaboration2021Special3}. Therefore, we adopt our calculated ages in subsequent analyses. We modify our procedure for the close binary V4046 Sgr. While our derived age for this system is in agreement with a portion of the published literature, it is not in agreement with the age of the $\beta$ Pictoris moving group for which it is a member. Our calculated age from our isochrone fit is 5.62$^{+4.38}_{-2.46}$ Myr and the age of the $\beta$ Pictoris moving group is 23$\pm$3 Myr \citep{Mamajek2014}. Notably, our calculated mass of 0.90$^{+0.09}_{-0.13}$ M$_{\tt \sun}$ is in good agreement with the dynamical mass of 0.90$\pm$0.05 M$_{\tt \sun}$ from \cite{Rosenfeld2012}. Older system ages result in higher mass estimates; therefore, to be most conservative, we adopt the older age of the moving group in case the luminosity measurement was contaminated from the binary component, which was noted as a possibility in \cite{McCarthy2012}. Moreover, owing to the fact that all methods of estimating the ages of young stars (both the method that we employ and the methods employed in the literature for these systems) are rife with assumptions that bias the resulting ages,  we also present the analyses in Section~\ref{planetary_mass} and Section~\ref{visco} independent of system age in \ref{sec:age_indep} to show the effect of age estimation on our mass limits. 

\subsection{Determining Stellar L\texorpdfstring{$'$} { } Magnitudes} \label{stellar_L}
In the absence of detecting any previously unknown point sources, we seek to place constraints on the upper limits of the masses of planets that could be creating the substructures seen at other wavelengths in these disks. We can utilize our contrast limits in each of the gaps and cavities indicated by longer wavelength observations. Owing to our observing wavelength, we are more sensitive than millimeter observations to directly detecting the thermal emission of these proposed planets, and therefore our observations, despite being nondetections, can be useful in constraining the properties of these planets. 

Our contrast curves can be used to estimate our mass sensitivity when paired with a host star magnitude and an evolutionary model that predicts the expected planet mass from its magnitude. In order to estimate the stellar magnitude, we utilize the WISE (for systems with $W1<$4.5) and CATWISE (for systems with $W1>$4.5) $W1$ and $W2$ measurements for each of our systems (3.4 and 4.6 microns; \citealp{marocco21}), and determine a predicted $L'$ magnitude in order to determine the sensitivity of our observations. This is necessary as the WISE $W1$ and $W2$ bandpasses only cover a portion of the $L'$ bandpass. Following \cite{Keppler2018}, we estimate the corresponding $L'$ magnitude by interpolating logarithmically between the $W1$ and $W2$ bands. This is necessary because these protoplanetary disk-hosting stars show strong infrared excesses and therefore are often redder than would be expected for a bare stellar photosphere.

\subsection{Planetary Mass Upper Limits}\label{planetary_mass}
In order to determine the planetary masses corresponding to our contrast constraints, we use the AMES-Cond and AMES-Dusty models \citep{Allard2012, Baraffe2003Evolutionary209458}. The AMES-Cond and AMES-Dusty models provide the expected magnitude of a planet given a mass and age. We use these grids (interpolating between the grid points), the absolute magnitude limits  from our 5$\sigma$ contrast curves, and the estimated stellar ages to calculate the upper limits on the planet masses. An additional important caveat to our derived mass limits is that they are estimated given the model assumptions inherent in the AMES-Cond and AMES-Dusty models \citep{Cugno2023}. We show results using both sets of models to indicate a range of possible masses given two different model assumptions, and therefore a first-order indication of the sensitivity of our mass estimates on the chosen model.

Another important caveat for our derived masses is that we are not accounting for the effects of extinction from the circumstellar disk material. The circumstellar disk can extinct the light from a companion, meaning that our mass sensitivities may be overestimated. Indeed, when \cite{Cugno2023} accounted for the effects of 1.0 magnitudes of extinction, the sensitivity of their survey did decrease. However, the effects of that level of extinction decreased their sensitivity by less than a factor of 2, and the uncertainties on system ages and evolutionary models can affect the derived mass estimates by an even larger factor. Determining the magnitude of the extinction for each system is also a nontrivial task and comes with additional inherent uncertainties.  Therefore, while an important caveat to consider in our derived masses, we do not take extinction into account when presenting our mass limits.

High-resolution submillimeter observations of disks with substructures allow us to constrain the locations of possible gap-carving planets and can provide information about the masses of those possible planets. \cite{Zhang2018} carried out a series of 2D hydrodynamical simulations to infer the relationship between gaps in disks imaged with ALMA as part of the DSHARP survey \citep{Andrews2018} and planet mass. This leads to direct predictions of planet masses for the objects that could be clearing out the gaps in the millimeter observations of these systems. \cite{Wang2021} utilized ALMA disk morphology and estimated the masses of planetary substructure drivers using the pebble isolation mass for a sample of systems that included the DSHARP \citep{Andrews2018} sample and the systems imaged in \cite{Long2018}. \cite{lodato19} also estimated the masses of planets in systems imaged as part of the DSHARP survey \citep{Andrews2018} and systems in \cite{Long2018} using empirical scaling relationships. These different methods of estimating gap-opening planet masses result in differences in the masses of these objects, which can all be compared against our companion mass upper limits.

\begin{figure*}[htb!]
\begin{centering}
\includegraphics[width=.99\textwidth]{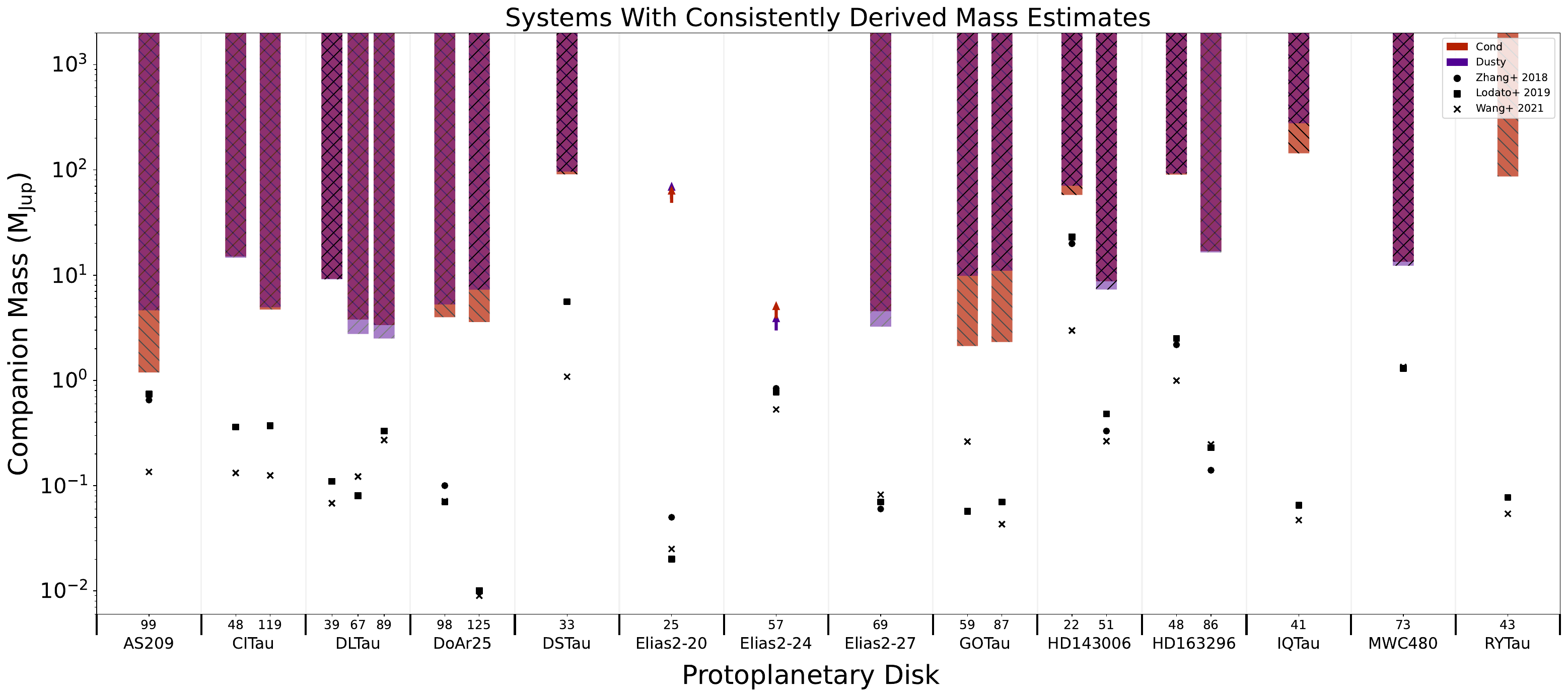}
\caption{We show a comparison of the mass estimates from the ALMA data and our observational mass limits for the subset of our systems observed as part of the DSHARP survey \citep{Andrews2018} or the survey of the Taurus molecular cloud \citep{Long2018}. We show the ALMA derived planet masses (in M$_{\rm Jup}$) as the points and the mass estimates derived using the AMES-Cond models from our 5$\sigma$ contrast limits as the orange bars for each observed gap from the ALMA data (specified in astronomical units as the numbers above each system designation). The bars therefore represent the range of masses that we would have been sensitive to given our observations. The ages for Elias~2-20 and Elias~2-24 are lower than the youngest age in the AMES-Cond grid (1~Myr). Therefore, we show the masses assuming 1~Myr as upper limits to the ages, as younger systems would result in smaller masses. We also show the associated masses assuming the AMES-Dusty models when the masses were within the grid in purple.} 
\label{figure:gaps}   
\end{centering}
\end{figure*}

For systems from \cite{Long2018} or \cite{Andrews2018}, Figure~\ref{figure:gaps} compares our mass limits, as computed from our observed 5$\sigma$ contrast limits at the radial location of the planet estimated from the ALMA radial profiles, to the masses predicted in \cite{lodato19}, \cite{Wang2021}, and \cite{Zhang2018}. We account for projection effects using published inclinations in each of the references cited in Table~\ref{table:systems} and account for the new parallaxes from Gaia \citep{GaiaCollaboration2021Special3} to adjust the radial separations when necessary. When systems are inclined, planetary orbits trace out ellipses, resulting in changing angular separations from the host star with inclination. In order to be the most conservative in our mass limits, we take the smallest angular separation consistent with the inclination. Overall, our current mass limits are not sensitive enough to probe down to the expected masses of the planets derived from the ALMA data, consistent with the lack of new point source detections herein.

\begin{figure*}[htb!]
\begin{centering}
\includegraphics[width=1\textwidth]{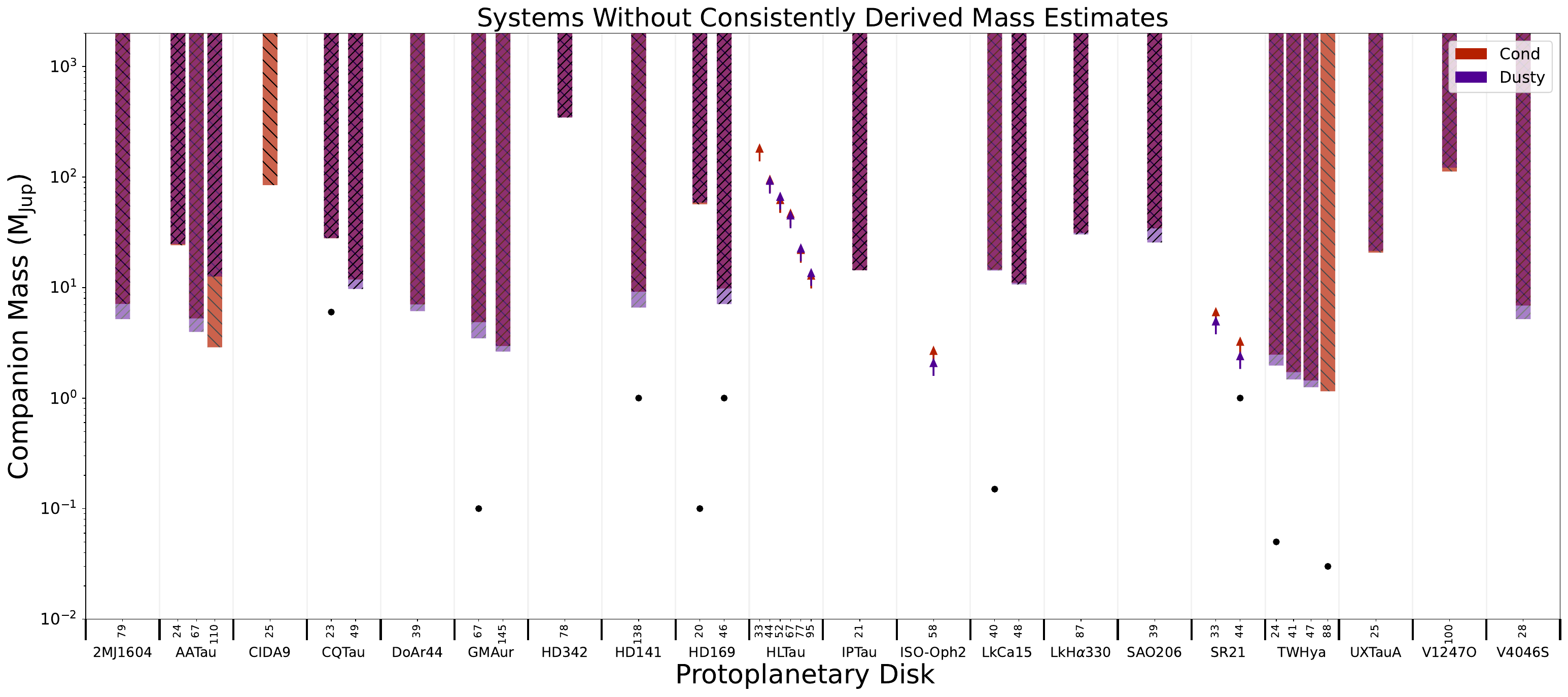} 
\caption{We show the mass estimates derived from our 5$\sigma$ contrast limits (in M\textsubscript{Jup}) in each gap/cavity (specified in astronomical units above each system designation) as the bars colored by radial distance from the host star. The bars therefore represent the range of masses that we would have been sensitive to given our observations. The ages for HL~Tau, ISO-Oph~2, and SR~21 are lower than the youngest age in the AMES-Cond grid (1~Myr), therefore we show the masses assuming an age of 1~Myr as a limit to the ages, as a younger system would correspond to smaller mass. We also show the associated masses assuming the AMES-Dusty models when the masses were within the grid in purple. There is a subset of systems that have estimates of planetary masses not ascertained via the \citet{lodato19}, \citet{Wang2021}, and \citet{Zhang2018} methods. The putative companion in CQ Tau is detailed in \citet{wolfer2021}, the putative companion in GM Aur is detailed in \citet{Huang2020ADisk}, and the putative companion in HD 141569 is detailed in \citet{Konishi2016}. The putative companions in each of the two gaps in HD 169142 have mass estimates of between 0.1-1M$\textsubscript{Jup}$ for the inner planet and 1-10M$\textsubscript{Jup}$ for the outer planet, so we show the lower limits for these masses \citep{Fedele2017}. The putative companion in SR 21 is detailed in \citet{Muro-Arena2020}. The putative companions in TW~Hya and LkCa 15 are detailed in \citet{Dong2017}. We show the expected planet masses as points. Additional details on NIRC2 observations of TW Hya with the vortex coronagraph can be found in \citet{Ruane2017}. For the locations of the central cavities, we generally use the inner edges of the dust rings.} 
\label{figure:cavities}   
\end{centering}
\end{figure*}

Many of our other observed systems have detected gaps in the millimeter continuum images but are not of high enough quality to be able to use the aforementioned methods from \citet{lodato19}, \citet{Wang2021}, and \citet{Zhang2018} to ascertain the planet mass. Some of these systems also show a large inner cavity devoid of dust at longer wavelengths. Mass estimation methods from the ALMA data require a gap surrounded by dust, meaning that we cannot simply estimate the mass of a planet that could be carving out an inner cavity. This inner cavity could also be due to the accretion of disk material onto the host star, making decoupling the influence of the possible planet on the disk in that inner region difficult. What is more, it is possible that the planets could visually (but not physically) co-locate with the scattered light from small dust grains outside of large cavities in our observations, making disk modeling a likely necessity to disentangle the two physical signals (e.g., PDS 70c in \citealt{Wang202070}; also see \citealt{Quiroz2022} for more details about the benefits of utilizing disk modeling for increased sensitivity to planets). Additionally, bright disk signals at similar radii to planets may bias both contrast curve and S/N estimates of point-like sources and due to the filtering of extended signals into point-like sources, the use of dedicated techniques (e.g., MAYONNAISE \citep{Pairet2021}, REXPACO \citep{Flasseur2021}) may be necessary to reliably image both point sources and extended signals. However, there are a number of other methods to estimate the masses of planets that could be creating the cavities and gaps seen in the millimeter. We show the estimated masses of the putative companions, where available in the published literature, in Figure~\ref{figure:cavities}. 

\begin{figure}[htb!]
\begin{centering}
\includegraphics[width=.4\textwidth]{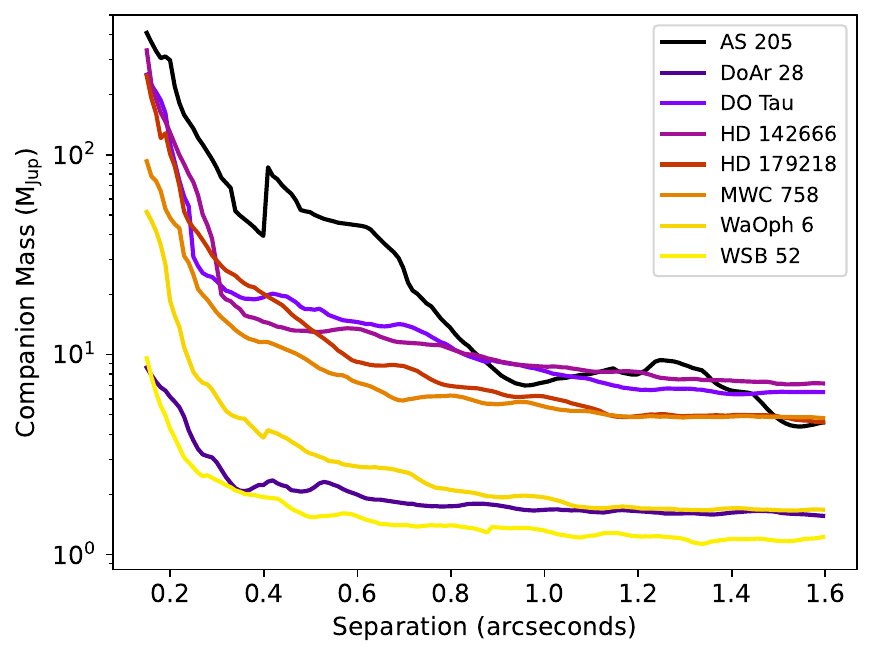} 
\caption{Companion mass upper limits (computed using the AMES-Cond model grid) as a function of the separation for the systems studied herein which did not have clear locations where a planet would be present or where the location of a gap/cavity was outside of our separation range. When the age of the system is lower than the youngest age in the AMES-Cond grid, we assume an upper limit of 1 Myr.} 
\label{figure:mass_with_sep}   
\end{centering}
\end{figure}

\begin{figure*}[htb!]
\begin{centering}
\includegraphics[width=\textwidth]{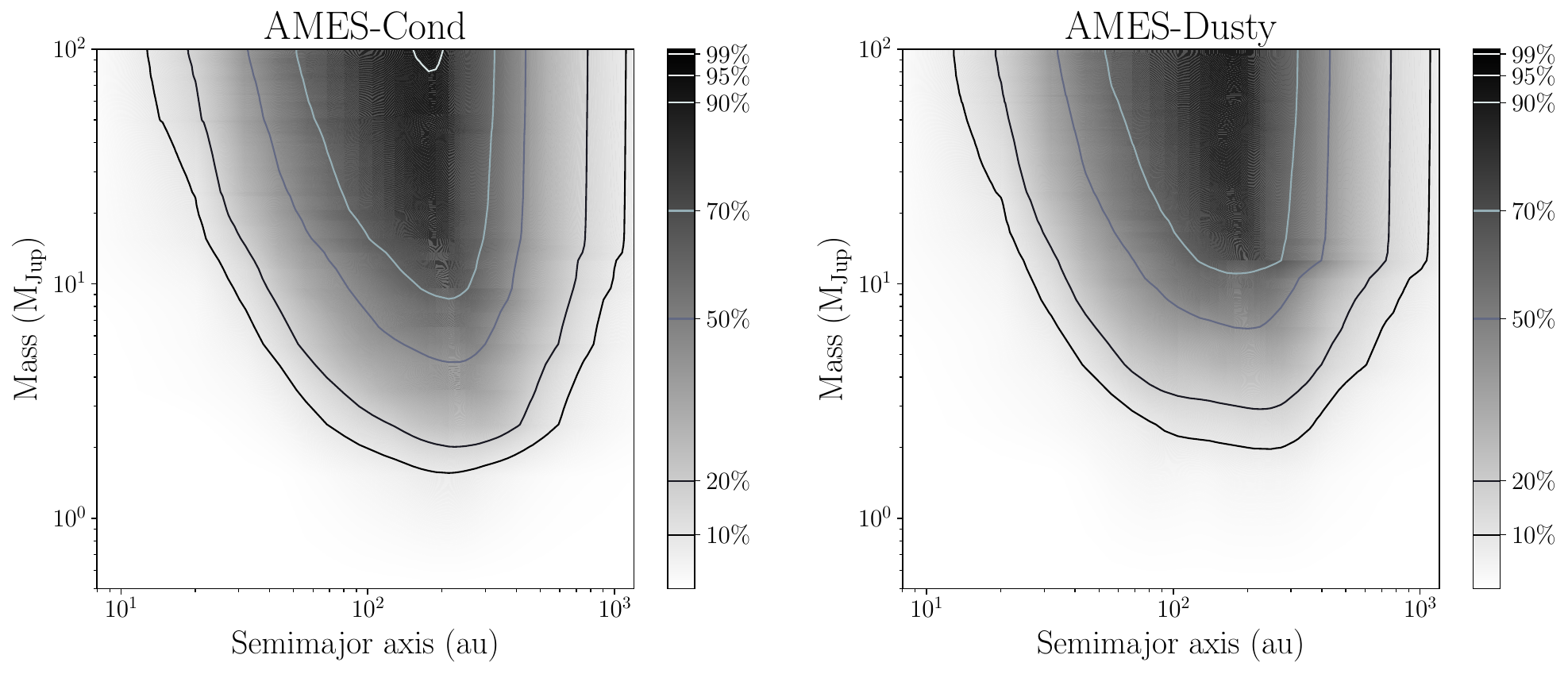} 
\caption{A map of the average sensitivity of our survey in planet mass (calculated using the AMES-Cond and AMES-Dusty models) vs. separation in astronomical units calculated using \texttt{Exo$-$DMC}  \citep{Bonavita2020}.} 
\label{figure:sensitivity}   
\end{centering}
\end{figure*}

The other systems in our survey do not have high enough resolution ALMA data (DoAr~28 and DO~Tau), did not have clear evidence for a gap in millimeter images or exact gap locations were not determined but were taken as part of the DSHARP survey (\citealt{Andrews2018}; WSB~52 and AS~205), are spiral systems without a consistent radial distance for which a gap is cleared (WaOph~6 and MWC~758) or have gaps/cavities outside of the distances probed with our observations (HD 142666 and HD 179218). Therefore, instead of comparing our mass detection limits to locations of interest in each disk, we calculate our mass limits over all radial separations from the host star (Figure~\ref{figure:mass_with_sep}). Additionally, we show our average survey sensitivity over all separations in Figure~\ref{figure:sensitivity} determined using \texttt{Exo$-$DMC} \citep{Bonavita2020}. In brief, \texttt{Exo$-$DMC} uses a Monte Carlo approach to determine the fraction of planets on Keplerian orbits that would have been recovered given an observational mass limit with separation. We account for the published inclinations for each of our disk systems when calculating the separations and use the default underlying eccentricity distribution from \texttt{Exo$-$DMC}.

\section{Discussion}\label{sec:Discussion}
\subsection{Investigating Disk Viscosity Constraints} \label{visco}
Determining the mass of a planet carving out a gap at submillimeter wavelengths requires assumptions about the natal disk's viscosity, $\alpha$ \citep{Wang2021,Zhang2018}. Therefore, if we compare the calculated masses of planets using different values for the disk viscosity, and find that our sensitivity at $L'$ predicts a mass that is below one or more of the masses calculated using different disk viscosities, we may be able to place observational constraints on this parameter with the caveat that we are assuming the AMES-Cond and AMES-Dusty models. In order to compare our mass limits with those of masses determined with different values for $\alpha$, we again utilize the subset of our observations with ALMA data taken as part of the DSHARP survey \citep{Andrews2018} or the survey of systems in the Taurus molecular cloud \citep{Long2018}. The DSHARP survey is one of the highest angular resolution surveys at millimeter wavelengths, which allows for robust planetary mass predictions. Moreover, \cite{Zhang2018} directly computed the predicted masses of substructure drivers for different values of $\alpha$. \cite{Wang2021} also directly computed the predicted masses of substructure driving planets at different values of $\alpha$ for systems in \cite{Long2018} and \cite{Andrews2018}.

\begin{figure*}[htb!]
\begin{centering}
\includegraphics[width=\textwidth]{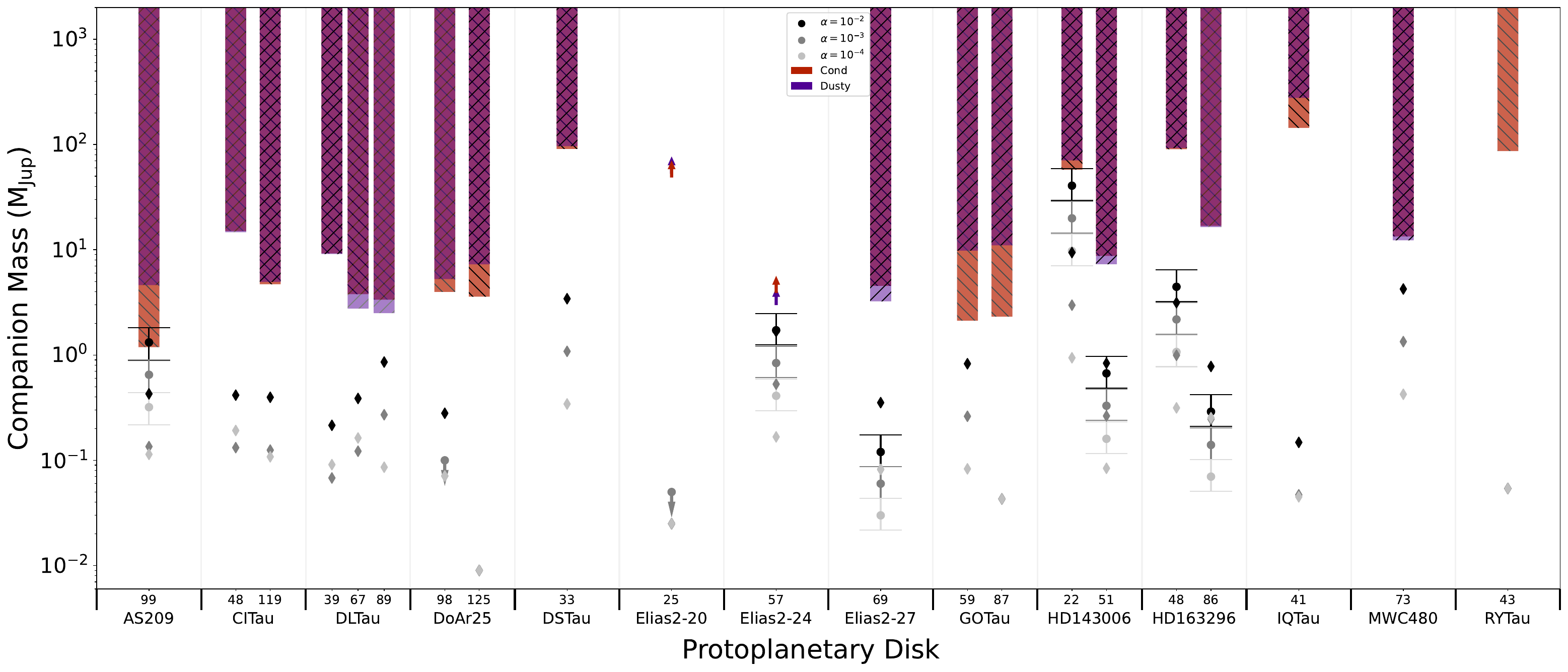}
\caption{Expected planet masses from \citet{Wang2021} (diamonds with no associated errors as none were reported) and \citet{Zhang2018} (points with their associated errors), using different values for the disk viscosity, $\alpha$, are shown as black and gray points. We show our planet mass detection limits for comparison using the AMES-Cond models and AMES-Dusty models (when the masses were within the AMES-Dusty grid) for the subset of our systems observed as part of the DSHARP survey \citep{Andrews2018} or the survey of the Taurus molecular cloud \citep{Long2018}. The bars therefore represent the range of masses that we would have been sensitive to given our observations. The ages of Elias~2-20 and Elias~2-24 are lower than the youngest age in the grids (1~Myr). Therefore, we show the masses assuming 1~Myr as upper limits to the ages, as younger systems would result in smaller masses.}
\label{figure:alpha}   
\end{centering}
\end{figure*} 

As shown in Figure~\ref{figure:alpha}, our $L'$ mass sensitivities do overlap with the masses predicted by the method in \citet{Zhang2018} for a number of targets when assuming $\alpha$=10$^{-2}$. These overlapping values within their errors means that we cannot conclusively exclude this $\alpha$=10$^{-2}$ value observationally, but it does indicate that we would possibly be sensitive to massive planets in AS~209, Elias~2-24 (the actual  limit on the mass is smaller than the limit indicated by the arrow since the age of the system is younger than the lowest age in the model grids), and HD~143006. As with the mass estimations, uncertainties in the ages estimated for the host stars also add uncertainty to conclusions that can be made about the disk viscosity (see Figure~\ref{figure:alpha_ages} for a comparison of the masses derived from our 5$\sigma$ contrast limits at different system ages with the masses estimated using different disk viscosities). The fact that we do not detect planets in any of these systems, might be tentative evidence that the disk viscosity for these systems (and indeed protoplanetary disks in general), might be less than this 10$^{-2}$ value. This is consistent with $\alpha$ constraints from millimeter CO measurements (for example, \citealp{Flaherty2018TurbulenceDisk} constrained the disk viscosity to $\alpha$ $<$0.007 within a narrow region around the midplane of TW Hya and \citealp{Villenave2022} found that $\alpha$ $\lesssim$10$^{-5}$ at 100 au in Oph 163131). However, an important caveat to this is that we are generally not sensitive enough to detect planets regardless of the disk viscosity assumed using the mass estimates from \cite{Wang2021}.

\subsection{Limits on Planetary Accretion Rates} \label{mdot}
\begin{figure*}[htb!]
\begin{centering}
\includegraphics[width=\textwidth]{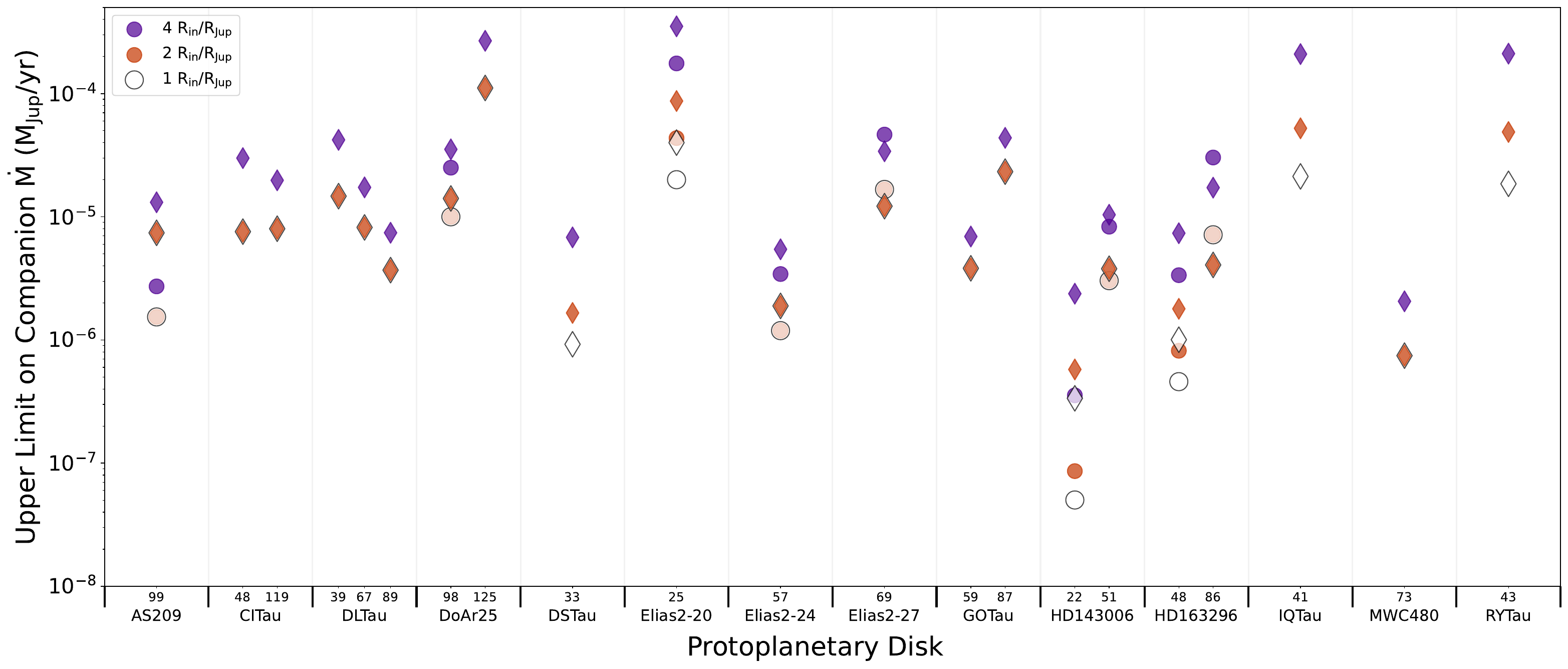}
\caption{Upper limits on the mass accretion rates as a function of circumplanetary disk radius, R$_{\rm in}$, for the subset of our systems observed as part of the DSHARP survey \citep{Andrews2018} or the survey of the Taurus molecular cloud \citep{Long2018}, derived from \cite{Zhu2015} and determined from our contrast limits and masses from \cite{Wang2021} (diamonds) and \cite{Zhang2018} (circles) assuming a disk viscosity of $\alpha$=10$^{-3}$.} 
\label{figure:mdot}
\end{centering}
\end{figure*}

Although we do not detect any new planets, and indeed do not seem to be able to achieve the sensitivity needed to directly detect the thermal emission of the planets thought to be creating the substructures seen at millimeter wavelengths (Figure ~\ref{figure:gaps}), we can still place observational constraints on the nature of these planets. As forming protoplanets accrete their envelopes from their natal disks, the brightness associated with this accretion can surpass the intrinsic luminosity of the planet's residual heat of formation \citep{Szulagyi2019,Zhu2015}. Therefore, assuming that we are not sensitive enough to detect the planet's thermal emission, we can utilize our contrast limits to place constraints on the accretion, assuming all of the luminosity that would be visible would be due to accretion. \citet{Zhu2015} related the expected magnitude to the circumplanetary accretion rate $M_{p}\dot{M}$. Generally, the mass of the planet cannot be disentangled from the mass accretion rate using infrared photometry, but utilizing the mass constraints from ALMA observations, and our contrast limits, we can break this degeneracy. We show the mass accretion rates for the planets studied herein with masses derived in \citet{Wang2021} and \citet{Zhang2018} as a function of circumplanetary disk radius, R\textsubscript{in}, in Figure~\ref{figure:mdot}. The upper limits to the mass accretion rates that we derive are larger than those from H$\alpha$ measurements of the actively accreting protoplanet PDS 70c (10$^{-8.0\pm0.4}$ M\textsubscript{Jup}yr$^{-1}$; \citealt{Benisty2021}, \citealt{Haffert2019}), but are consistent with those derived from infrared observational limits \citep{Ruane2017}.

\section{Conclusions}\label{sec:Conclusions}
We present new deep $L'$ observations of 43 protoplanetary disks using the NIRC2 vortex coronagraph at Keck. We primarily selected systems that had evidence of substructures in their millimeter/submillimeter continuum images in order to ascertain if we could directly detect planets that could be forming these substructures. While we do not detect any novel point sources, we are able to utilize our detection limits to place robust upper limits on the masses of planets in these disks. We present contrast curves for these systems and utilizing newly derived stellar ages, stellar $L'$ magnitudes, and the AMES-Cond and AMES-Dusty models \citep{Allard2012,Baraffe2003Evolutionary209458} are able to place upper limits on the masses of planets in these systems. While we are not sensitive enough to observe planets of the masses predicted for submillimeter observations, we do probe down to $\sim$1 M$_{\rm Jup}$ for a number of systems that we have observed. This means that if planets are the primary driver for the observed substructures in the submillimeter observations of these disks, then they must be under our observational sensitivity for each system and sub-Jovian for a number of our observed systems, given the caveat that we are not accounting for extinction and given the underlying assumptions of the AMES-Cond and AMES-Dusty models. Utilizing our mass upper limits, we also investigate constraints that we can place on the viscosity of the disk, possibly allowing us to exclude $\alpha>$10$^{-2}$. From our observational sensitivities and the masses derived from ALMA observations, we are able to place limits on the mass accretion rates for a subset of systems studied herein.

We also detect scattered light disks in a number of our observations, which would allow for a detailed study of the properties of the dust in protoplanetary disks when observed at multiple wavelengths. By virtue of the fact that we purposely selected targets that have been observed at other wavelengths, many of our disks that show evidence of structure have been observed in the near-infrared by facilities such as GPI or SPHERE. However, our additional observations in a different wavelength and/or utilizing the vortex coronagraph to probe smaller separations allows for a unique perspective into these systems. Multiwavelength studies of protoplanetary disks will provide insight into the radial distributions of different sized dust in these disks, ultimately allowing for a better understanding of the natal disk environment and the availability of planetary building blocks. We reserve extensive modeling and discussion of these disks to a forthcoming paper. 

Future observations of these systems with the next generation of larger primary mirror ground-based telescopes will likely be needed in order to reach down to the sensitivities required to detect these planets directly. Achieving the sensitivity needed to observe planets of the masses predicted by millimeter observations will be needed to definitively ascertain the nature of the disk substructures revealed by ALMA.
\clearpage
\begin{acknowledgements} 
This research is partially supported by NASA ROSES XRP, award 80NSSC19K0294. The Infrared Pyramid Wavefront Sensor at W. M. Keck Observatory was developed with support from the National Science Foundation under grants AST-1611623 and AST-1106391,
as well as the Heising Simons Foundation under the Keck Planet Imager and Characterizer project. The data presented herein were obtained at the W. M. Keck Observatory, which is operated as a scientific partnership among the California Institute of Technology, the University of California and the National Aeronautics and Space Administration. The Observatory was made possible by the generous financial support of the W. M. Keck Foundation. The authors wish to recognize and acknowledge the very significant cultural role and reverence that the summit of Maunakea has always had within the indigenous Hawaiian community. We are most fortunate to have the opportunity to conduct observations from this mountain. The computations presented here were conducted in the Resnick High Performance Center, a facility supported by the Resnick Sustainability Institute at the California Institute of Technology. Part of this work was carried out at the Jet Propulsion Laboratory, California Institute of Technology, under contract with NASA. This work has made use of data from the European Space Agency (ESA) mission {\it Gaia} (\url{https://www.cosmos.esa.int/gaia}), processed by the {\it Gaia} Data Processing and Analysis Consortium (DPAC, \url{https://www.cosmos.esa.int/web/gaia/dpac/consortium}). Funding for the DPAC has been provided by national institutions, in particular the institutions participating in the {\it Gaia} Multilateral Agreement. We thank the anonymous referee for their comments that improved the clarity of this work. Part of this research has received funding from the European Union's Horizon 2020 research and innovation programme under the Marie Sk\l odowska-Curie grant agreement No.~101103114. Part of this research has received funding from the European Research Council (ERC) under the European Union's FP7 and Horizon 2020 research and innovation programmes (grant agreements No 337569 and 819155), and from the Wallonia-Brussels Federation (grant for Concerted Research Actions). This project has received funding from the European Research Council (ERC) under the European Union’s Horizon Europe research and innovation program (grant agreement No. 101039651, project DiscEvol, PI G. Rosotti). MV research was supported by an appointment to the NASA Postdoctoral Program at the NASA Jet Propulsion Laboratory, administered by Universities Space Research Association under contract with NASA. KRS acknowledges support from the JPL Researchers on Campus Program. GG acknowledges the financial support of the Swiss National Science Foundation in the framework of the NCCR PlanetS. JK's research on disks orbiting nearby young stars is supported by NASA Exoplanets Program grant 80NSSC19K0292 to RIT. 
\end{acknowledgements} 

\begin{appendix}
\setcounter{table}{0}
\renewcommand{\thetable}{A\arabic{table}}
\clearpage

\section{System Parameters and Observing Details}
We show the system parameters in Table~\ref{table:systems} and the observation details in Table~\ref{table:observations}. 

\startlongtable
\begin{deluxetable*}{lllllllllll}
\tablecaption{System Parameters \label{table:systems}}
\tablehead{
System&Distance\tablenotemark{a}&T$_{\rm eff}$&L\tablenotemark{b}& M$_{\rm\star}$\tablenotemark{c}&Age\tablenotemark{c}&$L'$\tablenotemark{d}&Ref.\tablenotemark{e} \\
& (pc)&(K)&(L$_{\sun}$)&(M$_{\sun}$)& (Myr)& (mag)& }
\startdata
2MJ1604\tablenotemark{f}&145.31$\pm$0.57&4898$\pm$180&0.58$\pm$0.19&1.01$\pm$0.11&15.85$^{+12.33}_{-6.94}$&7.58$\pm$0.02&1,2\\
AA Tau&134.67$\pm$1.57&3763$\pm$173\tablenotemark{g}&0.41$\pm$0.10&0.56$^{+0.16}_{-0.12}$&2.51$^{+2.50}_{-1.10}$&7.19$\pm$0.02&3,4\\
AS 205&132.06$\pm$1.25&4266$\pm$295&2.28$\pm$0.75&0.88$^{+0.31}_{-0.26}$&0.79$^{+1.44}_{-0.44}$&4.19$\pm$0.34&5,6\\
AS 209&121.25$\pm$0.43&4266$\pm$295&1.42$\pm$0.67&0.87$^{+0.23}_{-0.22}$&1.58$^{+3.43}_{-0.95}$&6.53$\pm$0.03&5,5\\
CIDA  9&175.08$\pm$2.69&3585$\pm$165\tablenotemark{g}&0.21$\pm$0.05&0.45$^{+0.14}_{-0.10}$&4.47$^{+4.45}_{-2.23}$&8.91$\pm$0.01&3,7\\
CI Tau&160.32$\pm$0.53&4277$\pm$197\tablenotemark{g}&0.83$\pm$0.19&0.92$^{+0.14}_{-0.17}$&2.82$^{+2.19}_{-1.23}$&6.79$\pm$0.02&3,7\\
CQ Tau&149.37$\pm$1.34&6900$\pm$318&8.93$\pm$2.07&1.57$^{+0.11}_{-0.10}$&11.22$^{+11.17}_{-2.31}$&6.63$\pm$0.06&8,9\\
DL Tau&159.94$\pm$0.50&4277$\pm$197\tablenotemark{g}&0.65$\pm$0.15&0.92$^{+0.12}_{-0.15}$&3.98$^{+3.10}_{-1.74}$&6.78$\pm$0.02&3,7\\
DoAr 25&138.16$\pm$0.82&4266$\pm$295&0.96$\pm$0.46&0.86$^{+0.21}_{-0.20}$&2.51$^{+4.57}_{-1.39}$&7.24$\pm$0.02 &5,5\\
DoAr 28&135.60$\pm$0.49&4350$\pm$200&1.28$\pm$0.30&0.95$\pm$0.18&1.78$^{+1.38}_{-0.78}$&8.13$\pm$0.02&10,11\\
DoAr 44&146.32$\pm$0.49&4730$\pm$218&1.92$\pm$0.45&1.26$^{+0.12}_{-0.19}$&2.00$^{+1.55}_{-0.87}$&6.68$\pm$0.03&12,13\\
DO Tau&138.52$\pm$0.68&3806$\pm$175\tablenotemark{g}&0.22$\pm$0.05&0.61$^{+0.12}_{-0.13}$&7.08$^{+5.51}_{-3.10}$&5.98$\pm$0.05&3,14\\
DS Tau&158.35$\pm$0.53&3792$\pm$175\tablenotemark{g}&0.24$\pm$0.06&0.61$^{+0.12}_{-0.13}$&6.31$^{+4.91}_{-3.15}$&7.26$\pm$0.02&3,7\\
Elias 2-20&137.53$\pm$3.96&3890$\pm$269&2.22$\pm$1.07&0.58$^{+0.22}_{-0.13}$&0.63$^{+4.38}_{-0.41}$&5.95$\pm$0.05&5,5\\
Elias 2-24&139.26$\pm$1.24&4266$\pm$295&6.32$\pm$2.87&0.92$^{+0.36}_{-0.22}$&0.35$^{+1.23}_{-0.20}$&6.42$\pm$0.03&5,5\\
Elias 2-27&110.07$\pm$10.30&3890$\pm$269&0.82$\pm$0.51&0.61$^{+0.20}_{-0.18}$&2.51$^{+8.71}_{-1.80}$&7.22$\pm$0.02&5,5\\
GM Aur&158.11$\pm$1.22&4115$\pm$190\tablenotemark{g}&0.62$\pm$0.15&0.81$^{+0.14}_{-0.16}$&3.16$^{+2.46}_{-1.38}$&8.17$\pm$0.01&3,15\\
GO Tau&142.38$\pm$0.41&3516$\pm$162\tablenotemark{g}&0.21$\pm$0.05&0.40$^{+0.13}_{-0.10}$&3.16$^{+3.92}_{-1.38}$&8.84$\pm$0.02&3,7\\
HD 34282&308.61$\pm$2.20&9250$\pm$125&14.66$\pm$0.67&1.95$^{+0.02}_{-0.01}$&19.95$^{+15.53}_{-7.36}$&7.04$\pm$0.02&16,17\\
HD 141569&111.61$\pm$0.36&9750$\pm$125&25.34$\pm$0.58&2.20$\pm$0.01&17.78$^{+17.70}_{-8.87}$&6.70$\pm$0.02&16,18\\
HD 142666&146.25$\pm$0.46&7586$\pm$350&8.91$\pm$4.58&1.62$^{+0.11}_{-0.09}$&19.95$^{+15.53}_{-9.95}$&5.32$\pm$0.08&5,5\\
HD 143006&167.34$\pm$0.51&5623$\pm$259&3.91$\pm$1.34&1.68$^{+0.26}_{-0.25}$&4.47$^{+4.45}_{-1.95}$&6.28$\pm$0.04&5,5\\
HD 163296&100.97$\pm$0.42&9333$\pm$645&16.97$\pm$12.69&2.05$^{+0.16}_{-0.14}$&15.85$^{+15.77}_{-6.94}$&3.43$\pm$0.44&5,5\\
HD 169142&114.87$\pm$0.35&7500$\pm$346&5.90$\pm$5.60&1.58$^{+0.11}_{-0.09}$&19.95$^{+15.53}_{-8.73}$&6.30$\pm$0.04&19,20\\
HD 179218&260.09$\pm$2.23&9640$\pm$444&104.85$\pm$30.00&2.52$^{+0.34}_{-0.32}$&3.55$^{+14.23}_{-1.31}$&4.48$\pm$0.25&21,22\\
HL Tau&147.3$\pm$0.5\tablenotemark{h}&4400$\pm$203&9.25$\pm$5.50&1.03$^{+0.28}_{-0.20}$&0.32$^{+1.68}_{-0.17}$&5.07$\pm$0.08&23,24\\
IP Tau&129.38$\pm$0.29&3763$\pm$173\tablenotemark{g}&0.33$\pm$0.08&0.58$^{+0.14}_{-0.13}$&3.55$^{+3.53}_{-1.55}$&7.62$\pm$0.01&3,7\\
IQ Tau&131.51$\pm$0.62&3690$\pm$170\tablenotemark{g}&0.22$\pm$0.05&0.53$^{+0.14}_{-0.12}$&5.62$^{+5.60}_{-2.81}$&7.13$\pm$0.02&3,7\\
ISO-Oph 2&134.25$\pm$7.56&3467$\pm$160&0.71$\pm$0.17&0.31$^{+0.09}_{-0.05}$&0.40$^{+0.40}_{-0.24}$&8.44$\pm$0.02&25,26\\
LkCa 15&157.19$\pm$0.65&4277$\pm$197\tablenotemark{g}&0.77$\pm$0.18&0.92$^{+0.14}_{-0.16}$&3.16$^{+2.46}_{-1.38}$&7.67$\pm$0.02&3,27\\
LkH$\alpha$ 330&318.22$\pm$3.49&6220$\pm$287\tablenotemark{g}&16.55$\pm$3.84&2.24$^{+0.32}_{-0.26}$&3.16$^{+1.85}_{-1.17}$&6.40$\pm$0.04&3,12\\
MWC 480&156.22$\pm$1.26&8250$\pm$380&17.81$\pm$5.50&1.85$^{+0.12}_{-0.10}$&14.13$^{+17.50}_{-7.05}$&4.73$\pm$0.09&19,7\\
MWC 758&155.87$\pm$0.76&8130$\pm$375&16.30$\pm$3.79&1.84$^{+0.11}_{-0.09}$&12.59$^{+15.59}_{-5.51}$&5.18$\pm$0.10&28,29\\
PDS 70&112.39$\pm$0.24&3972$\pm$36&0.34$\pm$0.09&0.75$^{+0.03}_{-0.04}$&5.62$^{+4.38}_{-2.08}$&8.01$\pm$0.02&30,31\\
RY Tau&138.22$\pm$3.88&5930$\pm$273\tablenotemark{g}&11.93$\pm$2.77&2.28$^{+0.34}_{-0.31}$&2.51$^{+1.95}_{-1.10}$&3.82$\pm$0.11&3,7\\
SAO 206462&135.00$\pm$0.44&6250$\pm$125&5.17$\pm$0.12&1.45$^{+0.04}_{-0.01}$&10.00$^{+1.22}_{-1.09}$&5.26$\pm$0.09&16,32\\
SR 21&136.43$\pm$0.56&4571$\pm$211&3.77$\pm$0.88&1.20$^{+0.29}_{-0.25}$&0.71$^{+0.70}_{-0.31}$&6.04$\pm$0.03&25,33\\
TW Hya&60.14$\pm$0.05&3776$\pm$174\tablenotemark{g}&0.24$\pm$0.05&0.59$\pm$0.13&6.31$^{+4.91}_{-3.15}$&7.20$\pm$0.02&3,34\\
UX Tau A&142.23$\pm$0.67&4870$\pm$224\tablenotemark{g}&1.64$\pm$0.38&1.27$^{+0.09}_{-0.13}$&3.16$^{+3.15}_{-1.38}$&6.87$\pm$0.02&3,12\\
V1247 Ori&401.30$\pm$3.16&7250$\pm$334&14.59$\pm$1.50&1.80$^{+0.08}_{-0.06}$&7.08$^{+0.86}_{-0.77}$&6.42$\pm$0.03&35,36\\
V4046 Sgr&71.48$\pm$0.11&4260$\pm$196&0.49$\pm$0.11&0.90$^{+0.09}_{-0.13}$&5.62$^{+4.38}_{-2.46}$, 23$\pm$3\tablenotemark{i}&7.27$\pm$0.02&37,38\\
WaOph 6&122.53$\pm$0.35&4169$\pm$288&2.86$\pm$1.38&0.78$^{+0.28}_{-0.20}$&0.71$^{+2.45}_{-0.46}$&6.36$\pm$0.03&5,5\\
WSB 52&135.27$\pm$0.92&3715$\pm$257&0.70$\pm$0.34&0.51$^{+0.19}_{-0.17}$&1.78$^{+7.13}_{-1.22}$&7.32$\pm$0.02&5,5\\
\enddata
\tablecomments{
\tablenotetext{a}{We use parallax values from Gaia EDR3 \citep{GaiaCollaboration2021Special3}.}
\tablenotetext{b}{We scale the luminosities in the cited references to the new distances from Gaia EDR3 \citep{GaiaCollaboration2021Special3}.}
\tablenotetext{c}{Our ages and stellar masses, computed using fits to isochrones (as described in Section~\ref{stellar_parameters}).}
\tablenotetext{d}{We calculate $L'$ magnitudes by logarithmically interpolating the WISE $W1$ and $W2$ magnitudes. See Section \ref{stellar_L} for more information.} 
\tablenotetext{e}{References for the source of the initial stellar parameters (that were rescaled when necessary to account for new distances from Gaia EDR3 \citep{GaiaCollaboration2021Special3}) and for a different wavelength observation of the disk.}
\tablenotetext{f}{Full name: 2MASS J16042165-2130284}
\tablenotetext{g}{The T$_{\rm eff}$ for this system is determined using Table 5 in \citet{Herczeg2014}, where we interpolate between stellar types when necessary. }
\tablenotetext{h}{Parallax measurement from \citet{Galli2018TheRegion}.}
\tablenotetext{i}{While the mass that we derive from our isochrone fit is consistent with the published dynamical mass \citep{Rosenfeld2012} and our derived age (5.62$^{+4.38}_{-2.46}$ Myr) is consistent with a portion of the published ages for this system, the age is at odds with that of the $\beta$ Pictoris moving group (23$\pm$3 Myr) for which it is a member \citep{Mamajek2014}. V4046 Sgr is a tight binary, so despite the agreement of our masses, we elect to use 23$\pm$3 Myr as the age of this system to mitigate any possible contamination on the stellar luminosity from the binary component.}
}
\tablerefs{
 (1) \citet{Carpenter2014}, 
 (2) \citet{Zhang2014ComparisonJ160421.7-213028}, 
 (3) \citet{Herczeg2014}, 
 (4) \citet{Loomis2017},
 (5) \citet{Andrews2018}, 
 (6) \citet{Kurtovic2018},
 (7) \citet{Long2018}, 
 (8) \citet{Testi2003}, 
 (9) \citet{Gabellini2019},
 (10) \citet{Kim2013},
 (11) \citet{Rich2015NEAR-IRDISK},
 (12) \citet{Andrews2011ResolvedDisks},
 (13) \citet{Cieza2021}, 
 (14) \citet{Kwon2015RESOLVINGCARMA}, 
 (15) \citet{Huang2020ADisk}, 
 (16) \citet{Guzman-Diaz2021},
 (17) \citet{VanDerPlas2017An34282}, 
 (18) \citet{Konishi2016},
 (19) \citet{Meeus2012},
 (20) \citet{Fedele2017}, 
 (21) \citet{Menu2015},
 (22) \citet{Kluska2018ACavity}, 
 (23) \citet{vanderMarel2019},
 (24) \citet{Carrasco-Gonzalez2019},
 (25) \citet{Manara2015},
 (26) \citet{Gonzalez-Ruilova2020ACavity}, 
 (27) \citet{Isella2012On15}, 
 (28)\citet{boehler2018}, 
 (29) \citet{Dong2018},
 (30) \citet{Keppler2018},
 (31) \citet{Keppler2019HighlyALMA}, 
 (32) \citet{VanDerMarel2016},
 (33) \citet{Pinilla2015}, 
 (34) \citet{Nomura2016},
 (35) \citet{Kraus2013},
 (36) \citet{Kraus2017Dust-trappingOrionis}, 
 (37) \citet{McCarthy2012}, 
 (38) \citet{Rosenfeld2013TheSgr}}
\end{deluxetable*}

\clearpage
\startlongtable
\begin{deluxetable*}{ccccc}
\tablecaption{Observation Log\label{table:observations}}
\tablehead{
System   & UT Date & Total Number of Frames & Total Integration Time & Parallactic Angular rotation \\
   &  & & (s) & ($^\circ$)
}
\startdata
2MJ1604\tablenotemark{a}&2017-05-10&73&3285.0&36.2\\
AA Tau&2020-10-25&153&4590.0&161.3\\
AS 205&2019-06-18&43&1290.0&139.4\\
AS 209&2018-05-26&60&2700.0&33.0\\
&2018-05-28&145&4350.0&54.0\\
&2018-07-23&90&2700.0&27.1\\
&2018-07-30&120&3600.0&44.5\\
CIDA 9&2020-10-25&83&2490.0&18.5\\
CI Tau&2018-10-21&132&3960.0&161.2\\
&2018-12-23&87&2610.0&182.0\\
CQ Tau&2018-12-24&60&1800.0&111.0\\
DL Tau&2019-01-09&160&4800.0&220.3\\
DoAr 25&2019-06-08&64&1920.0&30.9\\
DoAr 28&2017-07-01&90&4050.0&41.7\\
DoAr 44&2017-06-13&54&1620.0&21.4\\
&2017-06-30&100&3750.0&42.1\\
DO Tau&2020-10-09&44&1188.0&19.0\\
DS Tau&2020-10-09&65&1950.0&78.9\\
Elias 2-20&2020-05-29&143&7150.0&72.4\\
Elias 2-24&2019-02-17&20&600.0&8.2\\
&2020-06-01&161&4830.0&56.9\\
Elias 2-27&2019-05-22&30&900.0&16.9\\
&2020-05-30&131&3930.0&54.7\\
GM Aur&2017-01-14&120&5400.0&126.7\\
GO Tau&2020-11-27&90&2700.0&11.5\\
HD 34282&2017-02-07&73&3285.0&61.3\\
&2017-10-12&49&1470.0&24.6\\
&2017-10-13&139&4170.0&60.7\\
HD 141569&2015-06-11&39&780.0&49.3\\
HD 142666&2019-06-18&75&2250.0&38.7\\
HD 143006&2019-05-21&76&2280.0&42.5\\
&2019-06-08&60&1800.0&29.2\\
HD 163296&2017-05-31&80&2400.0&40.4\\
HD 169142&2020-05-31&78&2340.0&28.9\\
HD 179218&2016-08-14&45&2025.0&130.1\\
&2016-09-11&25&1200.0&0.4\\
&2016-10-17&30&1440.0&0.8\\
&2017-06-01&80&2400.0&131\\
HL Tau&2015-10-22&10&250.0&0.1\\
&2017-12-26&105&2100.0&161.3\\
IP Tau&2017-10-03&130&5460.0&211.6\\
IQ Tau&2020-11-27&84&2520.0&16.5\\
ISO-Oph 2&2021-05-19&237&7110.0&68.4\\
LkCa 15&2015-10-22&81&2025.0&156.9\\
&2015-10-24&22&550.0&8.5\\
&2015-12-27&50&1500.0&129.6\\
&2017-01-12&101&2272.5&160.8\\
&2020-10-10&101&3030.0&43.7\\
LkH$\alpha$ 330&2015-10-22&44&1100.0&41.1\\
&2018-12-25&93&2790.0&38.3\\
MWC 480&2018-12-23&80&2400.0&132.2\\
&2018-12-24&140&4200.0&87.9\\
MWC 758&2015-10-24&81&2025.0&128.8\\
&2016-10-24&80&3200.0&173.2\\
PDS 70&2019-06-08&42&1260.0&27.8\\
RY Tau&2015-10-22&33&825.0&12.0\\
&2015-10-24&83&2075.0&105.3\\
SAO 206462&2016-05-27&80&2400.0&23.0\\
SR 21&2020-05-31&137&4110.0&56.3\\
TW Hya&2016-04-13&70&2100.0&21.6\\
&2017-01-09&120&5400.0&44.4\\
&2017-01-13&117&5265.0&44.7\\
&2019-02-17&64&1920.0&33.2\\
UX Tau A&2016-10-16&116&2320.0&199.5\\
&2017-02-11&50&1200.0&156.3\\
&2017-10-04&128&5376.0&164.2\\
V1247 Ori&2017-12-27&51&1530.0&4.0\\
V4046 Sgr&2017-05-11&70&3150.0&28.3\\
WaOph 6&2020-05-30&41&1230.0&13.3\\
WSB 52&2020-06-02&168&5040.0&68.3\\
\enddata
\tablenotetext{a}{Full name: 2MASS J16042165-2130284}
\end{deluxetable*}

\setcounter{table}{0}
\renewcommand{\thetable}{B\arabic{table}}
\section{Contrast Curves}
We show the most optimal 5$\sigma$ contrast curves for each target interpolated to selected separations in Table~\ref{table:contrasts}. 

\startlongtable
\begin{deluxetable*}{cccccccccc}
\tablecaption{5$\sigma$ Optimal Contrasts at Select Locations\label{table:contrasts}}
\tablehead{
	&	0$\farcs$15		&	0$\farcs$2	&	0$\farcs$4	&	0$\farcs$6	&0$\farcs$8	&	1$\farcs$0	&	1$\farcs$2	&	1$\farcs$4	&	 1$\farcs$6
}
\startdata
2M J16042165-213&3.60E-03&1.91E-03&3.01E-04&1.85E-04&1.09E-04&1.05E-04&9.29E-05&1.01E-04&8.29E-05\\
AA Tau&2.52E-03&1.03E-03&1.93E-04&1.40E-04&1.16E-04&1.05E-04&1.04E-04&1.03E-04&9.38E-05\\
AS 205&1.86E-02&1.35E-02&9.33E-04&9.07E-04&2.46E-04&1.11E-04&1.26E-04&9.62E-05&5.69E-05\\
AS 209&1.28E-03&6.35E-04&7.82E-05&4.84E-05&3.44E-05&2.86E-05&2.56E-05&2.49E-05&2.47E-05\\
CIDA 9&2.35E-02&9.21E-03&2.98E-03&2.23E-03&1.88E-03&2.54E-03&1.86E-03&1.70E-03&1.42E-03\\
CI Tau&2.44E-03&1.08E-03&2.15E-04&1.44E-04&1.16E-04&1.03E-04&9.51E-05&9.00E-05&8.72E-05\\
CQ Tau&1.36E-03&4.79E-04&6.93E-05&2.07E-05&1.23E-05&1.22E-05&1.06E-05&9.89E-06&8.90E-06\\
DL Tau&8.32E-04&3.27E-04&7.97E-05&4.69E-05&3.18E-05&3.74E-05&3.41E-05&3.07E-05&2.80E-05\\
DoAr 25&4.04E-03&1.46E-03&2.62E-04&1.74E-04&1.41E-04&1.21E-04&1.07E-04&1.00E-04&8.73E-05\\
DoAr 28&3.89E-03&2.37E-03&4.37E-04&3.49E-04&2.56E-04&2.36E-04&2.28E-04&2.20E-04&2.05E-04\\
DoAr 44&3.23E-03&1.26E-03&1.51E-04&9.14E-05&5.95E-05&6.69E-05&4.91E-05&5.96E-05&4.55E-05\\
DO Tau&1.47E-02&6.16E-03&7.80E-04&4.02E-04&2.37E-04&1.51E-04&1.00E-04&9.09E-05&9.48E-05\\
DS Tau&6.42E-03&2.70E-03&4.75E-04&2.76E-04&2.19E-04&1.65E-04&1.71E-04&1.62E-04&1.50E-04\\
Elias 2-20&3.03E-03&6.77E-04&1.55E-04&9.55E-05&5.77E-05&5.17E-05&4.84E-05&4.73E-05&4.61E-05\\
Elias 2-24&3.63E-03&1.38E-03&2.75E-04&1.80E-04&8.96E-05&6.14E-05&5.77E-05&4.97E-05&4.64E-05\\
Elias 2-27&6.11E-03&2.22E-03&6.80E-04&2.76E-04&1.90E-04&1.58E-04&1.38E-04&1.41E-04&1.28E-04\\
GM Aur&3.03E-03&1.52E-03&3.43E-04&2.96E-04&2.34E-04&2.11E-04&2.21E-04&2.01E-04&1.83E-04\\
GO Tau&8.00E-04&4.97E-04&3.74E-04&4.12E-04&3.67E-04&3.51E-04&3.81E-04&3.75E-04&4.18E-04\\
HD 34282&3.63E-03&1.70E-03&3.21E-04&1.75E-04&9.72E-05&1.04E-04&8.10E-05&8.57E-05&6.86E-05\\
HD 141569&7.87E-03&3.38E-03&9.09E-04&3.06E-04&2.10E-04&1.77E-04&1.60E-04&1.60E-04&1.59E-04\\
HD 142666&4.96E-03&1.50E-03&1.30E-04&9.24E-05&4.32E-05&2.77E-05&2.44E-05&1.97E-05&1.84E-05\\
HD 143006&2.46E-03&9.53E-04&1.40E-04&8.01E-05&6.29E-05&4.39E-05&3.78E-05&3.52E-05&3.56E-05\\
HD 163296&6.07E-03&2.38E-03&2.38E-04&7.09E-05&3.49E-05&1.94E-05&1.23E-05&7.34E-06&6.06E-06\\
HD 169142&3.81E-03&1.20E-03&1.41E-04&9.24E-05&6.99E-05&6.28E-05&5.97E-05&5.85E-05&5.09E-05\\
HD 179218&1.67E-03&7.08E-04&7.35E-05&2.32E-05&1.50E-05&1.23E-05&8.20E-06&8.35E-06&7.17E-06\\
HL Tau&1.41E-02&4.68E-03&4.77E-04&1.61E-04&9.99E-05&7.17E-05&4.37E-05&3.20E-05&2.77E-05\\
IP Tau&2.25E-03&9.30E-04&2.10E-04&1.38E-04&1.15E-04&1.00E-04&1.06E-04&1.10E-04&8.56E-05\\
IQ Tau&2.65E-02&6.73E-03&6.95E-04&4.21E-04&1.95E-04&1.51E-04&1.33E-04&1.37E-04&1.39E-04\\
ISO-Oph 2&6.43E-03&3.73E-03&6.62E-04&4.43E-04&3.23E-04&3.25E-04&3.75E-04&3.58E-04&5.34E-04\\
LkCa 15&3.04E-03&1.49E-03&4.07E-04&1.97E-04&1.63E-04&1.46E-04&1.28E-04&1.30E-04&1.20E-04\\
LkH$\alpha$ 330&2.66E-03&1.01E-03&1.18E-04&4.97E-05&3.22E-05&2.72E-05&2.40E-05&2.71E-05&2.50E-05\\
MWC 480&1.04E-03&4.40E-04&4.79E-05&1.71E-05&8.85E-06&8.39E-06&7.13E-06&6.85E-06&6.66E-06\\
MWC 758&1.15E-03&5.76E-04&6.41E-05&2.55E-05&1.91E-05&1.47E-05&1.13E-05&1.15E-05&1.10E-05\\
PDS 70&6.67E-03&3.68E-03&7.36E-04&4.67E-04&3.78E-04&3.20E-04&3.07E-04&2.82E-04&2.80E-04\\
RY Tau&1.30E-03&3.25E-04&7.29E-05&2.04E-05&1.35E-05&8.88E-06&6.59E-06&6.13E-06&5.52E-06\\
SAO 206462&8.51E-03&3.80E-03&1.95E-04&1.26E-04&6.15E-05&3.20E-05&2.14E-05&2.36E-05&1.96E-05\\
SR 21&1.51E-03&5.02E-04&6.45E-05&5.80E-05&3.96E-05&2.52E-05&2.34E-05&2.31E-05&2.02E-05\\
TW Hya&2.01E-03&1.07E-03&2.50E-04&1.44E-04&7.59E-05&6.27E-05&4.96E-05&5.16E-05&4.61E-05\\
UX Tau A&2.44E-03&1.76E-03&2.25E-04&1.16E-04&8.59E-05&8.66E-05&8.29E-05&7.20E-05&6.35E-05\\
V1247 Ori&1.23E-03&6.96E-04&7.18E-05&2.87E-05&1.82E-05&1.75E-05&1.55E-05&1.41E-05&1.43E-05\\
V4046 Sgr&2.99E-03&1.46E-03&2.12E-04&1.93E-04&1.05E-04&9.95E-05&7.37E-05&8.99E-05&7.40E-05\\
WaOph 6&1.43E-02&3.62E-03&4.40E-04&2.48E-04&1.61E-04&1.32E-04&1.02E-04&1.01E-04&9.83E-05\\
WSB 52&2.14E-03&6.33E-04&1.56E-04&9.66E-05&7.97E-05&7.20E-05&6.49E-05&6.20E-05&6.42E-05\\
\enddata
\end{deluxetable*}

\section{Locations of Point Source Detections}
\begin{figure*}[h!]
{\includegraphics[width=\textwidth]{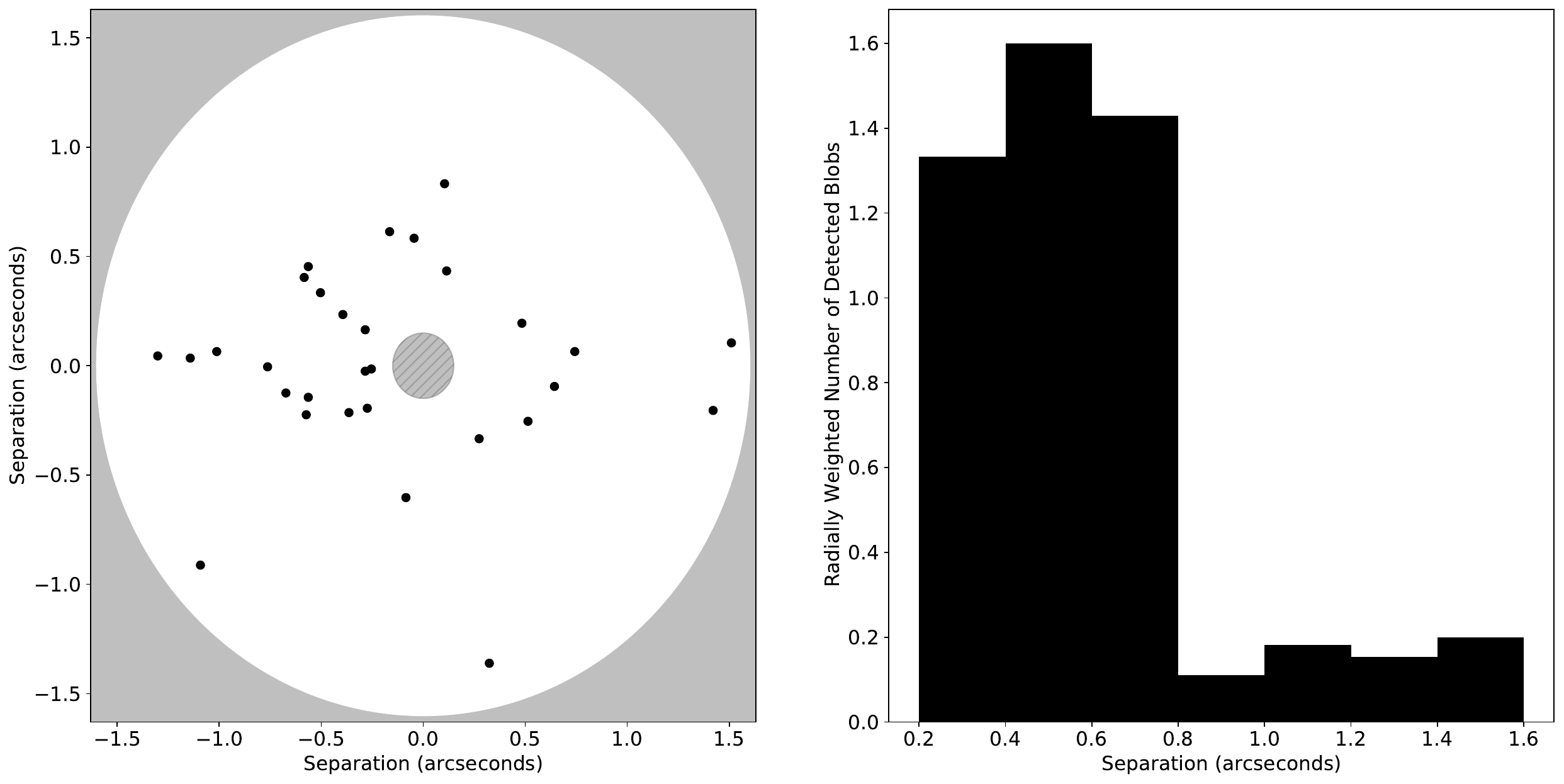}}
{\caption {Left: Locations of all point sources detected in our observations with S/N $>$5. The central $0\farcs15$ and separations$>$ $1\farcs6$ are hatched and represent the regions where we do not search for point sources. The clustering of points in radial lines is likely due to diffraction from the telescope support structure. Right: A histogram, weighted by area, of the separations of the detected point sources.}
\label{figure:blobs}}
\end{figure*}

We show the locations of the point sources that had S/N$>$5 between $0\farcs15$ and $1\farcs6$ in our \texttt{VIP} reductions. We vet these point sources in Section~\ref{point_sources} and conclude that none of them present convincing evidence that they represent astrophysical point sources. As evident by the weighted histogram in Figure~\ref{figure:blobs}, we have more false positives at smaller separations due to imperfect starlight suppression, and because of the higher false-positive rate for fixed S/N at smaller separations.

\section{Age Independent Mass Estimates}\label{sec:age_indep}
The ages of young stars are notoriously difficult to ascertain. In Figures~\ref{figure:gaps}-\ref{figure:mdot}, we utilize our best isochrone determined ages when estimating our planetary mass limits. Here we show our gap and cavity mass plots for a variety of system ages.

\begin{figure*}[htb!]
\begin{centering}
\includegraphics[width=\textwidth]{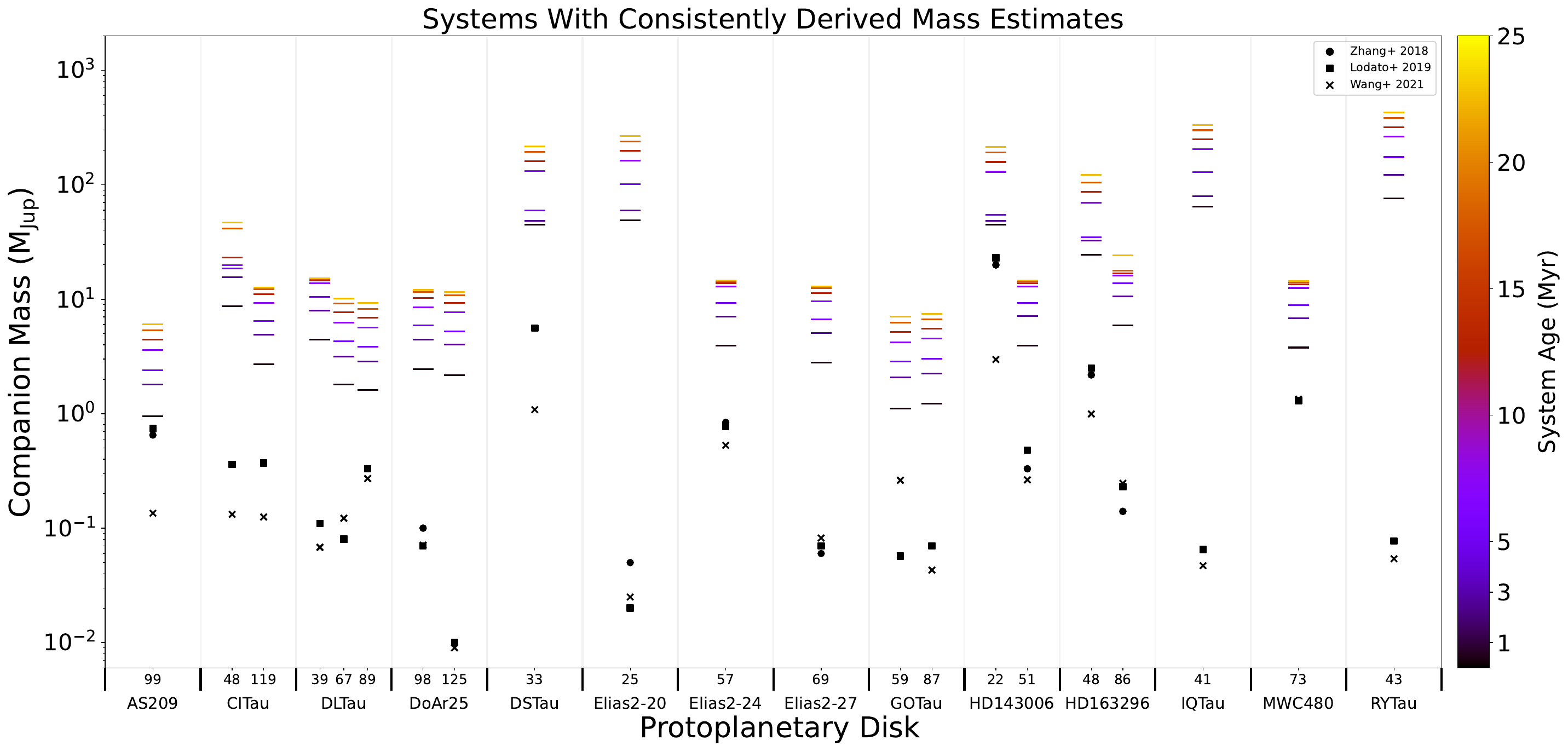}
\caption{Similar to Figure~\ref{figure:gaps}, but showing the mass estimates derived from our 5$\sigma$ contrast curves for a number of different system ages for systems observed as part of the DSHARP survey \citep{Andrews2018} or the survey of the Taurus molecular cloud \citep{Long2018}. We show the ALMA derived planet masses (in M$_{\rm Jup}$) as the points with our mass estimates using the AMES-Cond model shown as bars colored by system age. We specify the radial locations of the gaps that we are probing as numbers (in au) above each system name.} 
\label{figure:gaps_age}   
\end{centering}
\end{figure*} 

\begin{figure*}[htb!]
\begin{centering}
\includegraphics[width=\textwidth]{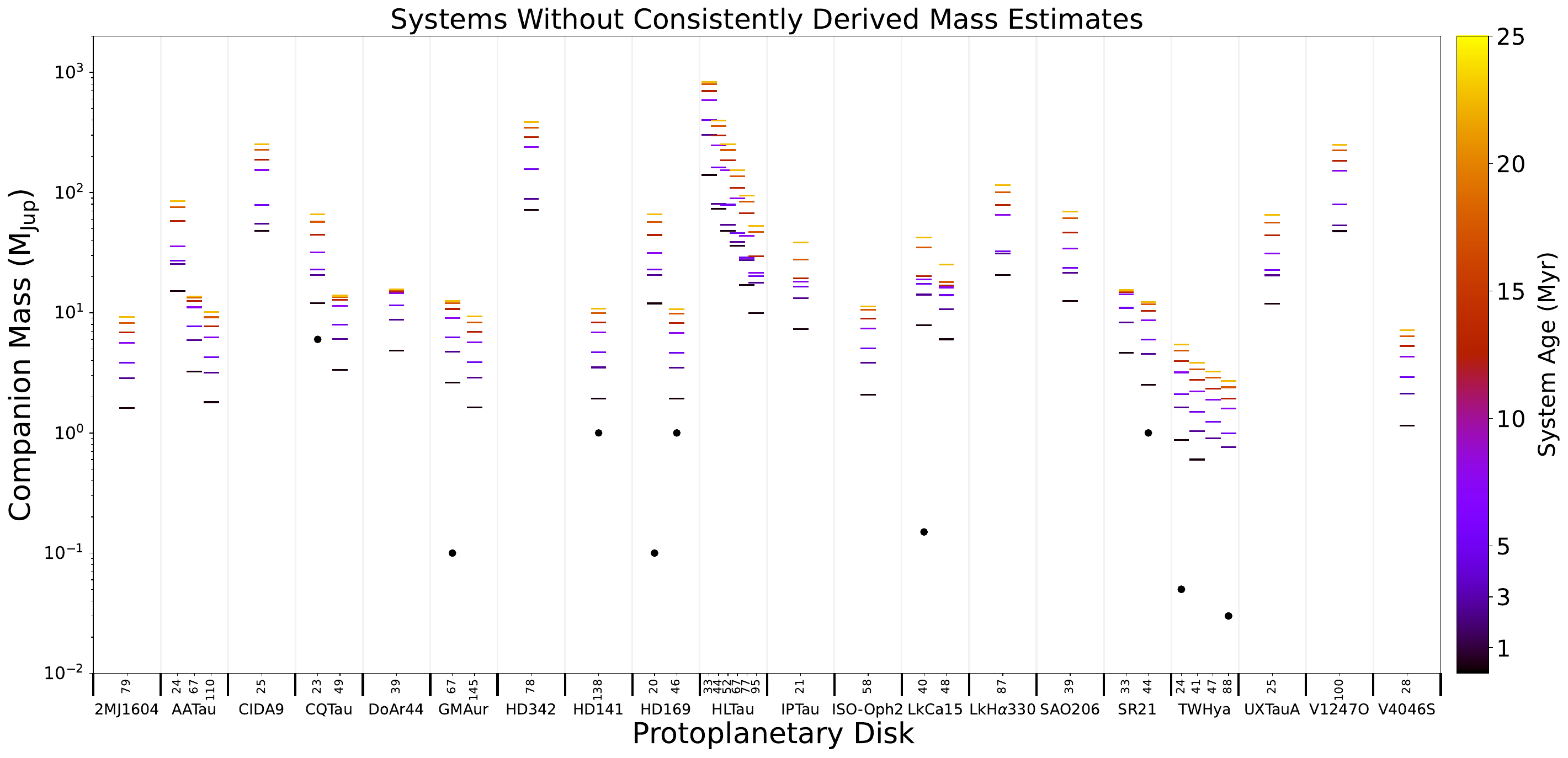}
\caption{Similar to Figure~\ref{figure:cavities}, but showing the mass estimates derived from our 5$\sigma$ contrast curves for a number of different system ages. We show our mass estimates using the AMES-Cond model as bars colored by system age. We specify the radial locations of the gaps that we are probing as numbers (in au) above each system name.} 
\label{figure:cavities_age}   
\end{centering}
\end{figure*}

\begin{figure*}[htb!]
\begin{centering}
\includegraphics[width=\textwidth]{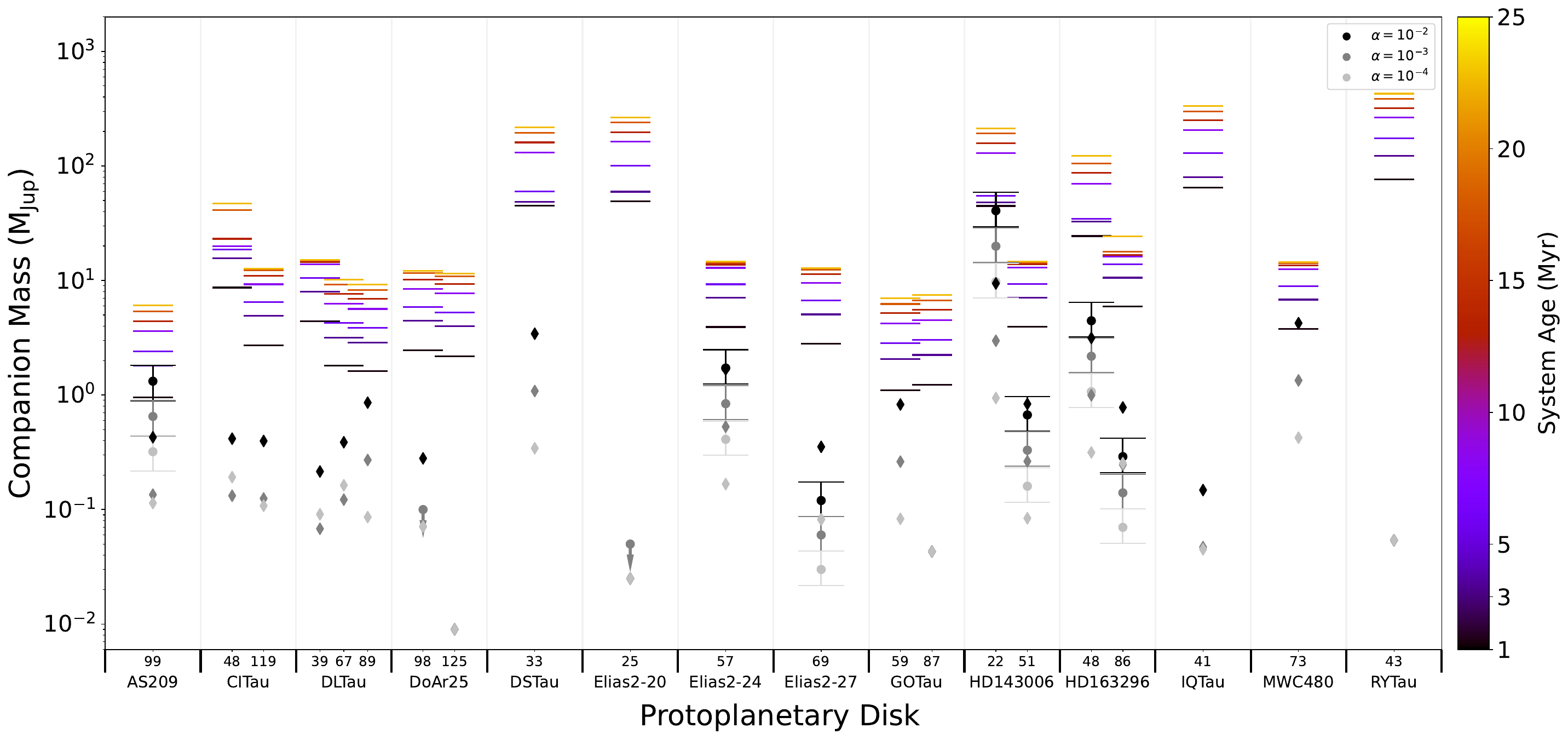}
\caption{Similar to Figure~\ref{figure:alpha}, expected planet masses from \citet{Wang2021} (diamonds with no associated errors as none were reported) and \citet{Zhang2018} (points with their associated errors) for the subset of our systems observed as part of the DSHARP survey \citep{Andrews2018} or the survey of the Taurus molecular cloud \citep{Long2018} calculated using different values for the disk viscosity, $\alpha$, are shown as black and gray points. We show the planet mass estimates derived from our 5$\sigma$ contrast limits for comparison using the AMES-Cond models \citep{Baraffe2003Evolutionary209458}.} 
\label{figure:alpha_ages}   
\end{centering}
\end{figure*} 
\end{appendix}
\clearpage
\bibliography{references.bib}
\end{CJK*}
\end{document}